\documentclass[a4paper, 11pt, onecolumn, accepted=2019-11-08]{quantumarticle}
\pdfoutput=1
\usepackage[english]{babel}
\usepackage[T1]{fontenc}
\usepackage{amsmath}
\usepackage{hyperref}

\usepackage[ansinew]{inputenc}
\usepackage{graphicx}
\usepackage{amsthm}
\usepackage{bm}
\usepackage{layout}
\usepackage{float}
\usepackage{amsfonts}
\usepackage{amssymb}
\usepackage{txfonts}%
\newcommand\id{\leavevmode\hbox{\small1\kern-3.3pt\normalsize1}}

\newcommand{\Tr}{\mbox{Tr}}

\begin{document}

\title{Time-delocalized quantum subsystems and operations: on the existence of processes with indefinite causal structure in quantum mechanics}

\author{Ognyan Oreshkov}

\affiliation{
QuIC, Ecole polytechnique de Bruxelles, C.P. 165, Universit\'e libre de Bruxelles, 1050 Brussels, Belgium}

\begin{abstract}

It has been shown that it is theoretically possible for there to exist higher-order quantum processes in which the operations performed by separate parties cannot be ascribed a definite causal order. Some of these processes are believed to have a physical realization in standard quantum mechanics via coherent control of the times of the operations. A prominent example is the quantum SWITCH, which was recently demonstrated experimentally. However, the interpretation of such experiments as realizations of a process with indefinite causal structure as opposed to some form of simulation of such a process has remained controversial. Where exactly are the local operations of the parties in such an experiment? On what spaces do they act given that their times are indefinite? Can we probe them directly rather than assume what they ought to be based on heuristic considerations? How can we reconcile the claim that these operations really take place, each once as required, with the fact that the structure of the presumed process implies that they cannot be part of any acyclic circuit? Here, I offer a precise answer to these questions: the input and output systems of the operations in such a process are generally nontrivial subsystems of Hilbert spaces that are tensor products of Hilbert spaces associated with systems at different times---a fact that is directly experimentally verifiable. With respect to these time-delocalized subsystems, the structure of the process is one of a circuit with a causal cycle. This provides a rigorous sense in which processes with indefinite causal structure can be said to exist within the known quantum mechanics. I also identify a whole class of isometric processes, of which the quantum SWITCH is a special case, that admit a physical realization on time-delocalized subsystems. These results unveil a novel structure within quantum mechanics, which may have important implications for physics and information processing.

\end{abstract}

\maketitle

\section{{Introduction}}\label{section1}

According to quantum mechanics, physical quantities in general do not have definite values unless measured. Yet, the classical idea that events occur in a well defined causal order persists, since quantum operations are assumed to always take place in acyclic compositions that respect the causal structure of spacetime. A natural question is whether this definiteness in the causal order of operations is a fundamental physical restriction or an artifact of our formulation of quantum theory. Is it possible that, in suitable circumstances, the order of operations would be indefinite similarly to other physical variables? How would this be described formally, and what testable consequences would it entail? These questions may be particularly relevant for understanding physics in the regimes of quantum gravity, where the causal structure of spacetime is expected to be subject to quantum indefiniteness \cite{hardyqg1, hardyqg2}. 

A theoretical model that can be interpreted as describing a situation in which separate operations occur in a `superposition' of different orders was proposed by Chiribella \textit{et al.} in Ref. \cite{Chiribella12}. This model, called \textit{quantum SWITCH}, was conceived as a hypothetical computer program that takes as an input two black-box quantum gates and outputs a new gate that can be thought of as the result of applying sequentially the two gates in an order that depends coherently on the logical value of a qubit that could be prepared in a superposition. Mathematically, such a program is an example of a higher-order quantum transformation, or supermap \cite{Chiribella08, networks}. The authors pointed out that this transformation cannot be realized by using each of the input gates once in an acyclic circuit \cite{onDeutsch}, just as the classical version of the program cannot, but nevertheless one can conceive implementations based on physical circuits with movable wires that simulate the effect of the program, thereby making a case for the need of a more general theoretical model of quantum computation than the circuit model. Although at face value the program involved nonclassicality in the order of operations, the theoretical tools for making this statement precise were not developed at that time. 

A framework for investigating the possibility of indefinite causal order in quantum theory by means of correlations was developed in Ref. \cite{OCB}. This so-called \textit{process framework} describes separate local experiments, each defined by a pair of input and output quantum systems on which an agent can apply arbitrary quantum operations, without presuming the existence of global causal order between the experiments. Under a set of natural assumptions (see Sec. \ref{section2}), the most general correlations between such experiments can be shown to be given by a generalization of Born's rule that involves an extension of the density matrix called the \textit{process matrix}. Mathematically, the process matrix can also be understood as a higher-order transformation, but one that maps the local operations to probabilities. The formalism provides a unified description of all nonsignaling and signaling quantum correlations between separate experiments that can be arranged in a causal configuration \cite{networks}, as well as probabilistic mixtures of different such causal scenarios. Remarkably, it was found that there are logically consistent bipartite process matrices that are incompatible with the existence of definite causal order between the local experiments and hence cannot be realized in this way. Such processes were called \textit{causally nonseparable} \cite{OCB}. The concept of causal nonseparability was subsequently generalized to more that two parties \cite{OG} (see also \cite{AraujoWitness, Wechs}), which in particular provided a rigorous framework in which the indefiniteness of the causal order involved in the quantum SWITCH can be defined. 

Causally nonseparable processes allow accomplishing certain tasks that cannot be achieved with operations for which a definite causal order exists. A striking possibility allowed by some causally nonseparable processes is the generation of correlations that violate \textit{causal inequalities} \cite{OCB, Baumeler1, Baumeler2, Baumeler3, OG, branciard,  Bhattacharya, Miklin, Feix2, Abbott}. These correlations imply incompatibility with definite causal order under theory independent assumptions, similarly to the way a violation of a Bell inequality implies incompatibility with local hidden variables \cite{Bell}. It is not known at present whether processes violating causal inequalities have a physical realization, except through post-selection \cite{OC2, Silvaetal, AraujoCTC, Milz}. However, it is widely believed that a specific class of causally nonseparable processes, which includes the quantum SWITCH, has a physical realization without post-selection via coherent control of the times at which the local operations occur, as in the original proposal \cite{Chiribella12}. The known processes of this kind cannot violate causal inequalities \cite{OG, AraujoWitness}, but they can be proven incompatible with definite causal order in a device-dependent fashion \cite{ Chiribella12b, AraujoWitness, BranciardWitness}. In particular, the quantum SWITCH and its generalizations have been shown to offer advantages over processes in which the order of operations is definite for a variety of information-processing tasks \cite{Chiribella12b, Colnaghi, Araujo, Feix, Guerin}. Concrete implementations of the quantum SWITCH via coherent control of the times of the operations have been proposed for trapped-ion systems \cite{Friis1}, photonic systems \cite{Procopio}, and systems of superconducting qubits \cite{Friis2}, and demonstrated with photonic systems in a series of increasingly sophisticated experiments \cite{Procopio, Rubino1, Rubino2, Goswami}. 

Despite this experimental progress, however, the question of whether implementations of this kind can be rightfully interpreted as realizations of causally nonseparable processes as opposed to some form of simulations of such processes has remained a subject of debate. The reason is that, even though by construction the produced data agrees with what we expect from the corresponding process, it is unclear whether the circumstances in which the data is produced correspond to a process. In a quantum process, by definition, the local operations of the parties occur each once on a specific pair of input and output Hilbert spaces. In an implementation based on controlled operations, it is not obvious on what spaces the local operations occur, if at all. For example, in the implementations of the quantum SWITCH \cite{Procopio, Rubino1, Rubino2, Goswami}, the experiment can be seen to involve two applications of controlled unitary operations at two different times in the laboratory of each party (see Sec. \ref{section3}), which are such that under the particular arrangement, when the control qubit is prepared in a classically definite logical state, exactly one of the two would result in a nontrivial transformation on the target system. However, in the actual implementation, the control qubit is prepared in a quantum superposition of the two logical states, in which case the belief that the same operation is applied once on the target system at some indefinite time is merely based on heuristics. This heuristics can be artificially strengthened by extending each controlled operation to act on a system that works as a `counter' that is coherently raised each `time' the nontrivial controlled operation is applied on the target system, such that the reading of the counter at the end could be interpreted as evidence that the desired operation has been applied once \cite{Araujo}. However, since the faithfulness of the counter as evidence for the applied operation is established only in the case when the target operation occurs at a definite time, regarding it as evidence in the case of superpositions requires the same conceptual leap. While such a leap appears appealing, it is nothing but begging for a rigorous theory that positions the supposed operations in relation to the standard temporal description of the experiment.

Here, I show that the supposed operations of the parties in such an implementation really take place on specific input and output systems---a fact that can be directly verified experimentally. These input and output systems are generally \textit{time-delocalized} subsystems, i.e., nontrivial subsystems of the tensor products of Hilbert spaces associated with different times. The fact that we can think, both mathematically and operationally, of Hilbert spaces that are tensor products of Hilbert spaces at different times is well established \cite{networks}---a generic fragment of a standard quantum circuit is an example of a quantum operation whose input and output Hilbert spaces are of this kind. However, given two Hilbert spaces of this kind, it is generally not possible to apply arbitrary quantum operations from one to the other due to the constraints imposed by the causal structure of spacetime \cite{networks}. At the same time, it is well known that the most general faithful realization of a quantum system inside a given Hilbert space is in the form of a subsystem---a tensor factor of a subspace of the full Hilbert space \cite{Viola, Knill, KribsSpekkens}. In fact, any experimentally accessible quantum system is such a subsystem from the perspective of a larger Hilbert space \cite{Zanardi, ZLL}. Thus, it is also natural to consider operations whose input and output systems are nontrivial subsystems of the tensor products of Hilbert spaces associated with different times. As it turns out, there exist pairs of such time-delocalized input and output subsystems on which it is possible to apply any standard quantum operation without post-selection, despite the fact that the input system cannot be associated with a region of spacetime that is in the causal past of the output system. It is on such time-delocalized quantum subsystems that causally nonseparable processes are realized. With respect to these input and output systems, a causally nonseparable process has the structure of a circuit with a cycle that does not admit a decomposition into a probabilistic mixture of acyclic circuits or dynamical generalizations of such mixtures \cite{OG}. The existence of these irreducible cyclic structures within quantum mechanics is the main finding of this paper. 

The rest of the paper is organized as follows. In Sec. \ref{section2}, I review the basics of the process matrix framework. In Sec. \ref{section3}, I describe explicitly the main result in the case of the quantum SWITCH. In Sec. \ref{section4}, I discuss the sense in which the result is regarded as justifying the interpretation of the discussed implementations as `realizations' of the quantum SWITCH, as well as the question of the resources used in these implementations. In Sec. \ref{section5}, I outline how the result generalizes to arbitrary processes in which the operations are delocalized in time through controlled operations. In Sec. \ref{section6}, I show that all bipartite processes that obey a recently proposed unitary extension postulate \cite{AraujoPostulate}, together with their unitary extensions, of which the quantum SWITCH is an example, have a realization on suitably defined time-delocalized subsystems. In Sec. \ref{section7}, I define a class of isometric extensions of bipartite processes, which is strictly larger than the class of unitary extensions, and show that these processes also admit a realization on suitable time-delocalized subsystems. It remains an open question whether the class of bipartite processes that admit such extensions is larger than those admitting unitary extensions. In Sec. \ref{section8}, I discuss the results.

\section{{The process matrix framework}}\label{section2}

The quantum process framework \cite{OCB} describes separate local experiments, $X= A, B, C,...$, each defined by an input quantum system $X_{I}$ with Hilbert space $\mathcal{H}^{X_{I}}$ and an output quantum system $X_{O}$ with Hilbert space $\mathcal{H}^{X_{O}}$, where an agent can perform an arbitrary quantum operation from $X_{I}$ to $X_{O}$. A quantum operation is most generally described by a collection of completely positive (CP) and trace-nonincreasing maps $\{ \mathcal{M}_{i^X}^{X_{I}\rightarrow X_{O}}\}_{i^X\in O^X}$, $ \mathcal{M}_{i^X}^{X_{I}\rightarrow X_{O}}:  \mathcal{L}(\mathcal{H}^{X_{I}})\rightarrow \mathcal{L}(\mathcal{H}^{X_{O}})$, where $\mathcal{L}({\mathcal{H}^X})$ denotes the space of linear operators over the Hilbert space $\mathcal{H}^X$ with dimension $d_X$ (here we assume finite dimensions), and $i^X\in O^X$ labels the possible outcomes of the operation with which the different CP maps are associated. The sum of the CP maps corresponding to the complete set of the outcomes of an operation, $\overline{\mathcal{M}}^{X_{I}\rightarrow X_{O}} = \sum_{i^X\in O^X}  \mathcal{M}_{i^X}^{X_{I}\rightarrow X_{O}}$, must be a CP and trace-preserving (TP) map. 

Consider the joint probabilities $p(i^A, j^B, \cdots  |\{ \mathcal{M}_{i^A}^{A_{I}\rightarrow A_{O}}\}_{i^A\in O^X}, \{ \mathcal{M}_{j^B}^{B_{I}\rightarrow B_{O}}\}_{j^B\in O^B}, \cdots )$ for the outcomes of the local experiments, conditional on the parties choosing to perform specific quantum operations. Under the following assumptions \cite{OCB}---(i) these probabilities are functions only of the CP maps corresponding to the local outcomes, (ii) they are consistent with the local quantum description of coarse-graining and randomization, (iii) the local operations can be extended to act on arbitrary auxiliary input systems prepared in any joint quantum state---the probabilities can be written in the form
\begin{gather}
p(i^A, j^B, \cdots  |\{ \mathcal{M}_{i^A}^{A_{I}\rightarrow A_{O}}\}_{i^A\in O^A}, \{ \mathcal{N}_{j^B}^{B_{I}\rightarrow B_{O}}\}_{j^B\in O^B}, \cdots )\notag\\
= \textrm{Tr} \left[W^{A_{I}A_{O}B_{I}B_{O}\cdots}\left(M^{A_{I}A_{O}}_{i^A}\otimes N_{j^B}^{B_{I}B_{O}}\otimes \cdots\right)\right]. \label{Wmain}
\end{gather}
Here, $M_{i^A}^{A_{I}A_{O}} \in{\cal L}({\cal H}^{A_{I}}\otimes{\cal H}^{A_{O}})\geq 0$ is the Choi-Jamio{\l}kowski (CJ) operator \cite{jam, choi} of the CP map $\mathcal{M}_{i^A}^{A_{I}\rightarrow A_{O}}$, and similarly for $N_{i^B}^{B_{I}B_{O}}$, etc. [Here, the CJ operator $M^{A_{I}A_{O}}\in{\cal L}({\cal H}^{A_{I}}\otimes{\cal H}^{A_{O}})$ of a linear map ${\cal M}^{A_I\rightarrow A_O}:{\cal L}({\cal H}^{A_{I}})\rightarrow {\cal L}({\cal H}^{A_{O}})$ is defined as $M^{A_{I}A_{O}} :=\big[{\cal I}^{A_I\rightarrow A_I} \otimes{\cal M}^{A_I'\rightarrow A_O}\big(|\phi^+\rangle\langle \phi^+|^{A_IA_I'}\big)\big]^{\mathrm T}$, where $|\phi^+\ \rangle^{A_IA_I'}=\sum_{j=1}^{d_{A_{I}}}|j\rangle^{A_I}|j\rangle^{A_I'} \in {\cal H}^{A_{I}}\otimes{\cal H}^{A_{I}'}$ is a (not normalized) maximally entangled state on two copies of $\mathcal{H}^{A_{I}}$ (one of them denoted by $\mathcal{H}^{A_{I}'}$), the states $\left\{|j\rangle^{A_I}\right\}_{j=1}^{d_{A_{I}}}$ form an orthonormal basis of ${\cal H}^{A_{I}}$, and $\left\{|j\rangle^{A_I'}\right\}_{j=1}^{d_{A_{I}}}$ are their copies in ${\cal H}^{A_{I}'}$, ${\cal I}^{A_I\rightarrow A_I}$ is the identity superoperator on ${\cal L}({\cal H}^{A_{I}})$, and ${\mathrm T}$ denotes matrix transposition in the above basis of $A_{I}$ and some basis of $A_{O}$ (see Ref. \cite{OC2} for a physical interpretation of this isomorphism and the choice of basis, which is based on the symmetry transformation of time reversal \cite{OC1}).] The operator $W^{A_{I}A_{O}B_{I}B_{O}}\cdots $ is an operator on the tensor product of all input and output systems, called the \textit{process matrix}. The only constraints that a process matrix must satisfy come from the requirement that probabilities must be nonnegative and sum up to 1 (the latter is equivalent to requiring normalization on all deterministic local operations, or CPTP maps):
\begin{gather}
W^{A_{I}A_{O}B_{I}B_{O}\cdots}  \geq 0,
\end{gather}
\begin{gather}
\textrm{Tr} \left[W^{A_{I}A_{O}B_{I}B_{O}\cdots}\left({M}^{A_{I}A_{O}} \otimes N^{B_{I}B_{O}}\otimes \cdots\right)\right] = 1, \nonumber \\
\forall {M}^{A_{I}A_{O}}, N^{B_{I}B_{O}} \cdots \geq 0,\nonumber\\
\textrm{Tr}_{A_{O}}M^{A_{I}A_{O}}=\id^{A_{I}}, \textrm{Tr}_{B_{O}}N^{B_{I}B_{O}}=\id^{B_{I}},   \cdots .\label{cond1}
\end{gather}

Condition \eqref{cond1} can be equivalently formulated as a simple constraint on the types of nonzero terms permitted in the expansion of a process matrix in a Hilbert-Schmidt basis, which is highly useful for constructing process matrices or verifying if a given operator is a valid process matrix \cite{OCB}. More specifically, any Hermitian matrix  
$W^{A_{I}A_{O}B_{I}B_{O}C_{I}C_{O}\cdots}$ can be written in the form
\begin{gather}\nonumber
 W^{A_{I}A_{O}B_{I}B_{O}C_{I}C_{O}\cdots} = \sum_{i,j,k,l,m,n\cdots} w_{ijklmn\cdots} \sigma^{A_{I}}_i\otimes\sigma^{A_{O}}_j\otimes\sigma^{B_{I}}_k\otimes\sigma^{B_{O}}_l\otimes\sigma^{C_{I}}_m\otimes\sigma^{C_{O}}_n\otimes \cdots,\\
w_{ijklmn\cdots}\in \mathbb{R}, \hspace{0.2cm}\forall i,j,k,l,m,n,\cdots,\label{Wexpansion}
\end{gather}
where the Hermitian operators $\{\sigma_\mu^X\}_{\mu = 0}^{d_X^2 - 1}$, with $\sigma_0^X = \id^X$, $\Tr\sigma_\mu^X\sigma_\nu^X = d_X \delta_{\mu\nu}$, and $\Tr\sigma_j^X = 0$ for $j=1,...,d_X^2 - 1$, form a Hilbert-Schmidt basis. An operator of the form \eqref{Wexpansion} satisfies condition \eqref{cond1}, if and only if, in addition to the term proportional to $\id^{A_{I}A_{O}B_{I}B_{O}C_{I}C_{O}\cdots}$ which comes with weight $w_{000000\cdots} = \frac{1}{d_{A_{I}}d_{B_{I}} d_{C_{I}}\cdots}$, it contains only nonzero terms in which there is a nontrivial $\sigma$ (different from the identity) operator on $X_{I}$ and a trivial one (the identity) on $X_{O}$ for some party $X\in\{A,B,C, \cdots\}$ \cite{OG,AraujoWitness}.

\begin{figure}
\vspace{0.5 cm}
\begin{center}
\includegraphics[width=11.5cm]{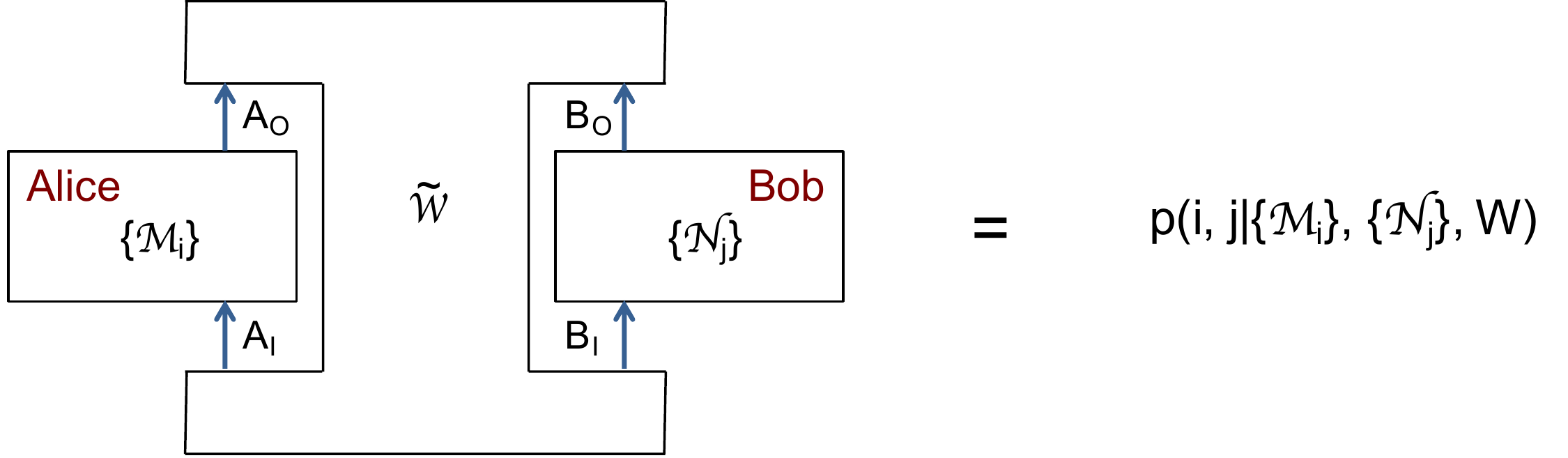}
\end{center}
\vspace{-0.5 cm}
\caption{\textbf{A process as a cyclic circuit.} A process matrix $W$ is equivalent to the transpose of the CJ operator of a channel $\widetilde{\mathcal{W}}$ from the outputs of the local operations to their inputs, while the probability rule \eqref{Wmain} is equivalent to composing this channel with the local operations, as illustrated here.} \label{fig:1}
\end{figure}

The process matrix can be understood as a higher-order transformation from the tensor product of a set of local quantum operations to conditional probability distributions, which is completely positive and normalized on deterministic local operations \cite{Perinotti, Kissinger}. Since a process matrix is mathematically equivalent to the transpose of the CJ operator of a channel $\widetilde{\mathcal{W}}^{A_{O}B_{O}\cdots \rightarrow A_{I}B_{I}\cdots}$ from the outputs of the local parties to their inputs (which can be seen from the terms permitted in its Hilbert-Schmidt basis expansion \cite{OCB}), and the probability formula (Eq.~\eqref{Wmain}) is equivalent to the composition of that channel with the local operations, such a higher-order transformation can be thought of as a circuit with a cycle \cite{OCB}, as illustrated in Fig. \ref{fig:1} for the case of two parties. Hereafter, when it is clear from the context, we will often refer to the channel $\widetilde{\mathcal{W}}^{A_{O}B_{O}\cdots \rightarrow A_{I}B_{I}\cdots}$ corresponding to a process matrix simply as the \textit{process}.  

In the case of two parties, the most general process matrix compatible with a fixed causal relation between the operations of Alice and Bob has the form $W^{A_{I}A_{O}B_{I}B_{O}} = W^{A_{I}A_{O}B_{I}}\otimes \id^{B_{O}} $ (the case where Bob is not in the causal past of Alice and hence cannot signal to Alice) or $W^{A_{I}A_{O}B_{I}B_{O}} = W^{A_{I}B_{I}B_{O}}\otimes \id^{A_{O}} $ (the case where Alice is not in the causal past of Bob and hence cannot signal to Bob). Each of these has an implementation via embedding the operations of Alice and Bob in an acyclic quantum circuit \cite{Chiribella08}. More generally, we may conceive of situations where these two possibilities are realized at random with some probabilities, in which case the process matrix can be written in the form 
\begin{gather}
W^{A_{I}A_{O}B_{I}B_{O}} _{cs} = q W^{A_{I}A_{O}B_{I}}\otimes \id^{B_{O}} + (1-q) W^{A_{I}B_{I}B_{O}}\otimes \id^{A_{O}}, \notag\\
q\in [0,1].\label{cs}
\end{gather}
Such process matrices are called causally separable \cite{OCB}. They correspond to the most general situation in which a fixed (though possibly unknown) casual configuration between the parties exist in each run of the experiment, and where the correlations arising in each configuration come from a standard quantum circuit containing the operations of Alice and Bob. (In the case of three or more parties, the concept of causal separability is more complicated due to the possibility of \textit{dynamical} causal configurations \cite{OG}, where the causal order between a subset of the parties can depend on the operations performed by other parties in their past.)

\section{{The quantum SWITCH and its physical realization}}\label{section3}

\begin{figure}
\vspace{0.5 cm}
\begin{center}
\includegraphics[width=12cm]{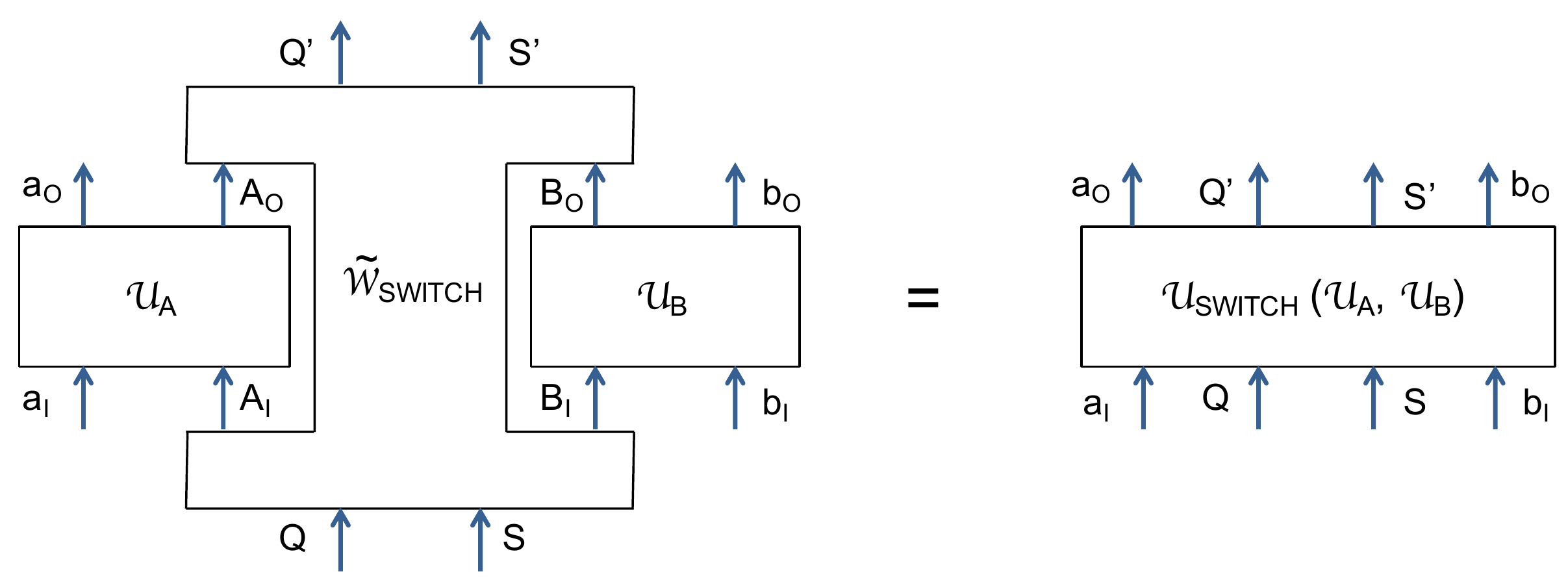}
\end{center}
\vspace{-0.5 cm}
\caption{\textbf{Extending local operations.} The quantum SWITCH, just like any higher-order process, is defined on all extensions of the local operations onto auxiliary systems. Hence, it is fully determined by its action on extended local unitary channels.} \label{fig:2}
\end{figure}

The quantum SWITCH \cite{Chiribella12} is a higher-order transformation, or supermap, which takes as an input two black-box operations, $\{ \mathcal{M}_{i^A}^{A_{I}\rightarrow A_{O}}\}_{i^A\in O^A}$ and $\{ \mathcal{N}_{j^B}^{B_{I}\rightarrow B_{O}}\}_{j^B\in O^B}$, figuratively associated with Alice and Bob, where $d_{A_{I}}= d_{A_{O}}=d_{B_{I}}=d_{B_{O}}=d$, and gives as an output an operation $\{ \mathcal{M}_{k}^{G\rightarrow G'}\}_{k\in O^A\times O^B}$, where $\mathcal{H}^{G} = \mathcal{H}^{Q} \otimes \mathcal{H}^{S} $ and $\mathcal{H}^{G'} = \mathcal{H}^{Q'} \otimes \mathcal{H}^{S'} $ with $d_Q=d_{Q'}=2$ and $d_S=d_{S'}=d$. Its action can be described intuitively as follows. If we think of $\mathcal{H}^Q$ and $\mathcal{H}^{Q'}$ as the Hilbert spaces of a control qubit at some initial and some final time, respectively, and of $\mathcal{H}^S$ and $\mathcal{H}^{S'}$ as the Hilbert spaces of some target system at the same two times, then the effect of the resultant operation can be thought of as transforming the target system from the initial to the final time by the sequential application of the operations $\{ \mathcal{M}_{i^A}\}_{i^A\in O^A}, \{ \mathcal{N}_{j^B}\}_{j^B\in O^B}$, where the order in which the two operations are applied depends \textit{coherently} on the logical value of the control qubit. To describe this coherent conditioning precisely, we use the fact that supermaps are defined on all extensions of the original input operations onto additional systems \cite{Chiribella12, Chiribella08, networks, Perinotti, Kissinger}. Since any quantum operation can be realized by applying a joint unitary channel on the original input system plus a suitably initialized auxiliary input system followed by a destructive measurement on a subsystem of the output system of the channel, the effect of the quantum SWITCH on the operations of Alice and Bob can be inferred from its effect in the case when the two operations are extended unitary channels $\mathcal{U}_A^{a_{I}A_{I}\rightarrow a_{O}A_{O}}$ and $\mathcal{U}_B^{b_{I}B_{I}\rightarrow b_{O}B_{O}}$, each acting on the original input and output systems plus separate auxiliary input and output systems, as described in Fig.~\ref{fig:2}. Let ${U}_A^{a_{I}A_{I}\rightarrow a_{O}A_{O}}$ and ${U}_B^{b_{I}B_{I}\rightarrow b_{O}B_{O}}$ denote the unitary matrices describing the action of these unitary channels at the Hilbert-space level. Then, the result of the quantum SWITCH is a unitary channel $\mathcal{U}_{\textrm{SWITCH}}^{a_{I}b_{I}G\rightarrow a_{O}b_{O}G'} {(\mathcal{U}_A, \mathcal{U}_B )}$ whose unitary matrix at the Hilbert-space level is 
\begin{gather}
U_{\textrm{SWITCH}}^{a_{I}b_{I}G\rightarrow a_{O}b_{O}G' }{(U_A, U_B )}\equiv U_{\textrm{SWITCH}}^{a_{I}b_{I}QS\rightarrow a_{O}b_{O}Q'S' }{(U_A, U_B )}=\nonumber\\
|0\rangle^{Q'} \langle 0|^Q \otimes     U_B^{b_{I}X\rightarrow b_{O}S'}\circ U_A^{a_{I}S\rightarrow a_{O}X}  \notag\\
+ |1\rangle^{Q'} \langle 1|^Q \otimes   U_A^{a_{I}X\rightarrow a_{O}S'}\circ U_B^{b_{I}S\rightarrow b_{O}X}, \label{ResultSwitch}
\end{gather}
where $X$ is a dummy system of dimension $d$ over which the transformations are composed (we have dropped the superscripts indicating the input and output systems of Alice's and Bob's operations in the argument of the supermap to simplify the notation). 

\begin{figure}
\vspace{0.5 cm}
\begin{center}
\includegraphics[width=13.5cm]{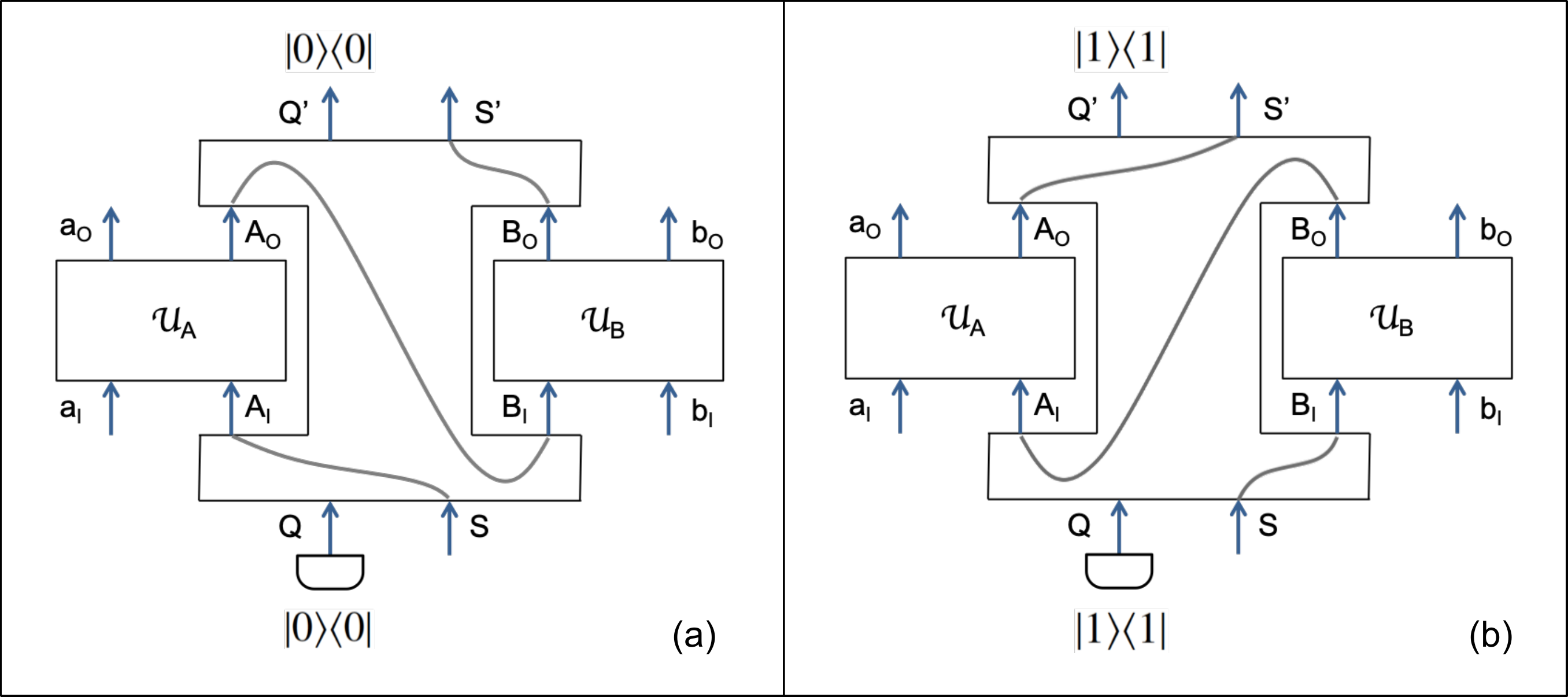}
\end{center}
\vspace{-0.5 cm}
\caption{\textbf{The quantum SWITCH in the extreme classical cases.} When the control qubit is initialized in either the state $|0\rangle\langle 0|$ [case (a)] or in the state $|1\rangle\langle 1|$ [case (b)], the result of the quantum SWITCH is effectively that of applying the operations of Alice and Bob on the target system in a particular order.} \label{fig:3}
\end{figure}

\begin{figure}
\vspace{0.5 cm}
\begin{center}
\includegraphics[width=6.5cm]{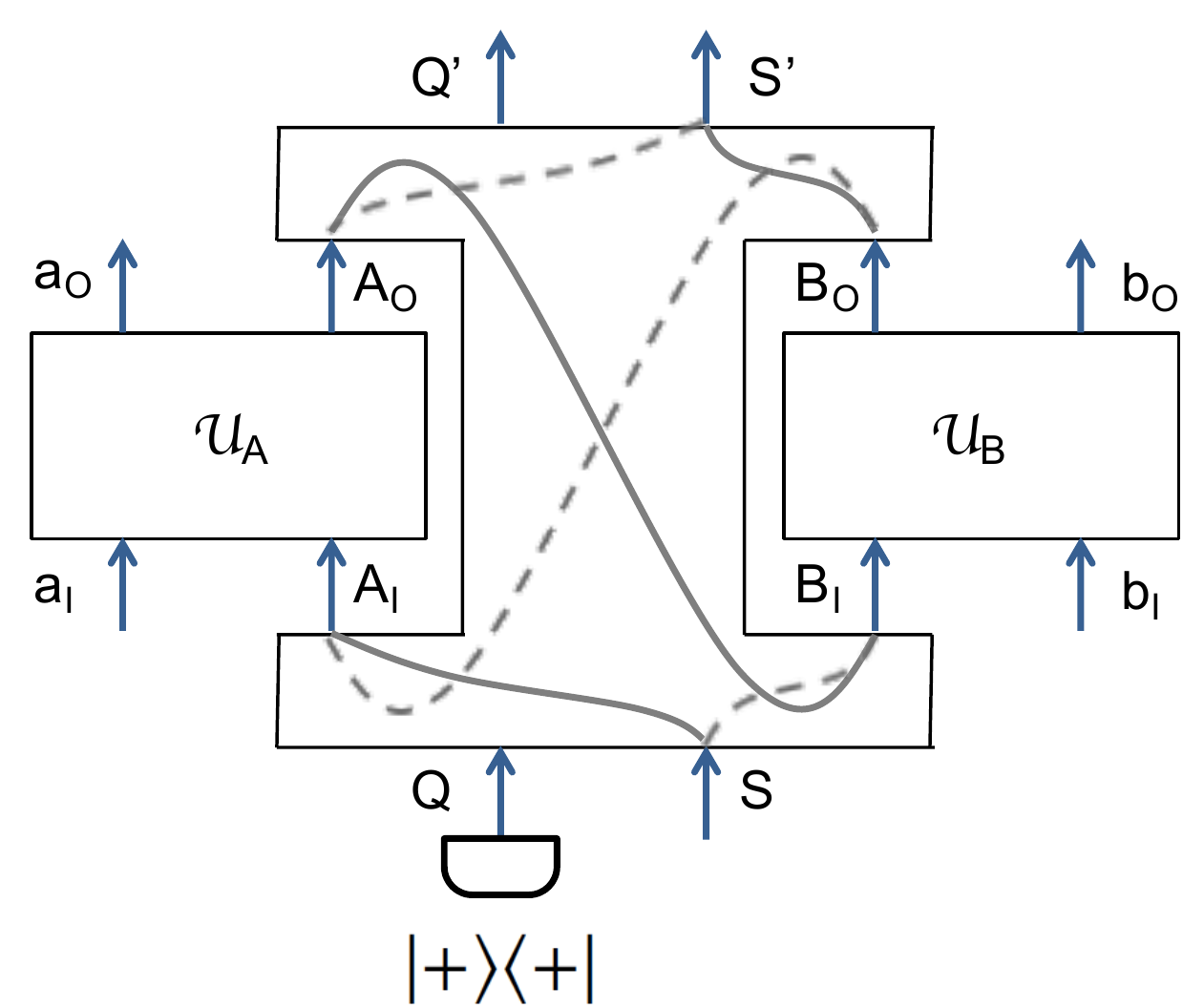}
\end{center}
\vspace{-0.5 cm}
\caption{\textbf{Intuitive description of the quantum SWITCH in the case of superpositions.} When the control qubit is initialized in a state that is a superposition of the logical basis states, such as $|+\rangle\langle +|$, one may intuitively think of the quantum SWITCH as realizing the extreme classical cases `in superposition'.} \label{fig:4}
\end{figure}

In the special cases when the control qubit is prepared in the state $|0\rangle\langle 0|^Q$, the quantum SWITCH effectively applies first the operation of Alice and then the operation of Bob on the target system. When the control qubit is prepared in the state $|1\rangle\langle 1|^Q$, it effectively applies  first the operation of Bob and then the operation of Alice on the target system (see Fig. \ref{fig:3}). When the control qubit is prepared in a superposition of these basis states, such as $|+\rangle\langle +|^Q$, $|+\rangle \equiv \frac{|0\rangle+|1\rangle}{\sqrt{2}}$, the intuitive understanding is that these two scenarios are somehow realized `in superposition', as sketched in Fig. \ref{fig:4}. Note, however, that the two extreme scenarios are not simultaneously compatible with a common causal structure \cite{Chiribella12}.

To see the quantum SWITCH as a process matrix, we consider two more parties---David, who is allowed to prepare different input states into the input system $G$ of the resultant channel (i.e., his possible operations have a trivial input system and output system $D_{O}\equiv G= QS$), and Charlie, who is allowed to perform measurements on the output system $G'$ of the resultant channel (i.e., his possible operations have the input system $C_{I}\equiv G'= Q'S'$ and a trivial output system). It can be verified that the process matrix describing the correlations between the four parties is 
\begin{gather}
W_{\textrm{SWITCH}}^{D_{O}A_{I}A_{O}B_{I}B_{O}C_{I}} = |W\rangle\langle W|^{D_{O}A_{I}A_{O}B_{I}B_{O}C_{I}},\label{nonsep01}
\end{gather}
where
\begin{gather}
|W\rangle^{D_{O}A_{I}A_{O}B_{I}B_{O}C_{I}} =   |0\rangle^{Q'} |0\rangle^{Q}  |\Phi^+\rangle^{A_{I}S}|\Phi^+\rangle ^{A_{O}B_{I}} |\Phi^+\rangle^{B_{O}S'} \notag\\
 +\   |1\rangle^{Q'}|1\rangle^{Q} |\Phi^+\rangle^{B_{I}S} |\Phi^+\rangle^{B_{O}A_{I}}|\Phi^+\rangle^{A_{O}S'},\label{nonsep02}
\end{gather}
with $ |\Phi^+\rangle  =\sum_i |i\rangle|i\rangle$, where $\{|i\rangle\}$ is the basis for the Choi isomorphism for the respective system.

The fact that this process matrix is causally nonseparable follows from the fact that there are certain preparations that David can make for which the operations of Alice and Bob cannot be said to take place in a definite order. The example most commonly considered and the one implemented in the experiments discussed below corresponds to the case where David prepares the state $|+\rangle\langle +|^Q$ on the control qubit together with some pure state $|\psi\rangle\langle \psi|^{S}$ on the target system. In that case, the correlations between Alice, Bob, and Charlie are given by the tripartite process matrix
\begin{gather}
W^{A_{I}A_{O}B_{I}B_{O}C_{I}} = |W\rangle\langle W|^{A_{I}A_{O}B_{I}B_{O}C_{I}},\label{nonsep1}
\end{gather}
where
\begin{gather}
|W\rangle^{A_{I}A_{O}B_{I}B_{O}C_{I}} = \frac{1}{\sqrt{2}} (    |0\rangle^{Q'}  |\psi\rangle^{A_{I}}|\Phi^+\rangle ^{A_{O}B_{I}} |\Phi^+\rangle^{B_{O}S'} \notag \\
 +  |1\rangle^{Q'}|\psi\rangle^{B_{I}} |\Phi^+\rangle^{B_{O}A_{I}}|\Phi^+\rangle^{A_{O}S'}).\label{nonsep2}
\end{gather}
The causal nonseparability of this process matrix \cite{OG, AraujoWitness} follows from the fact that it is proportional to a rank-one projector and hence cannot be written as a probabilistic mixture of different process matrices, but at the same time it permits some communication from Alice to Bob as well as from Bob to Alice---conditions that cannot be met simultaneously by a tripartite causally separable process matrix \cite{OG}. 

In the experimental implementations that we will discuss, the target system is represented by the internal degrees of freedom of a particle (e.g., the polarization of a photon as in Refs. \cite{Procopio, Rubino1, Rubino2}) and the control qubit by its path degree of freedom---an idea naturally inspired by Fig. \ref{fig:3} and Fig. \ref{fig:4}. (We stress, however, that our argument does not depend on the exact choice of physical encoding of the control qubit and the target system, and hence holds also for other realizations of this encoding, such as the one in Ref. \cite{Goswami}, where the control qubit is given by photon polarization and the target system by transverse spatial modes.) The setup is such that it allows the particle to follow two possible paths---along one path (corresponding to the initial state $|0\rangle\langle 0|^Q$) it first passes through Alice's laboratory and then through Bob's laboratory, and along the other path (corresponding to the initial state $|1\rangle\langle 1|^Q$) it passes through the laboratories in the opposite order. If inside each laboratory there is a device that applies the correct unitary on the internal degrees of freedom of the particle plus auxiliary systems when the particle passes through, it can be easily verified that the resulting transformation on all systems would be given by Eq. \eqref{ResultSwitch} (see Ref. \cite{Rubino1} for a concrete experimental implementation). 

The fact that the result of such an experiment is the transformation \eqref{ResultSwitch} does not by itself justify interpreting the experiment as a realization of the quantum SWITCH. The latter requires that this transformation be produced by applying each of the operations of Alice and Bob once on specific input and output systems. Whether this can be said to be the case in such an implementation is one of the central questions of this paper. The standard argument in favor of this being a realization of the process is merely based on the intuition that since the local devices would apply the correct operations on the internal degrees of freedom of the particle whenever the particles passes through at a given time, and since the particle may be heuristically argued to pass exactly once through each device (an idea that can be supported by supplying each device with a coherent counter that counts how many particles go through it), even if the time of passage in the two different classical cases is different and in the general case not even defined, we should be able to say that the correct operation has been applied once. While this line of reasoning may appear intuitive, the problem with it is that we only have a clear definition of what it means to apply the correct operations and possess evidence for this in the extreme cases of definite times. Simply declaring that this holds in the more general case without specifying where precisely the operations take place nor how we could probe them can hardly be considered convincing. (There are many properties that hold in the extreme case where a qubit is prepared in one of the two logical basis states but do not hold in the case of nontrivial superpositions---the causal separability of the presumed process resulting from such an implementation would be one example.) In fact, the above heuristic reasoning suggests that even in the case of indefinite times, the operations of Alice and Bob are still operations on the internal degrees of freedom of the particle. As we will see below, this is not correct---Alice and Bob can indeed be said to apply the correct operations, but the precise input and output systems are nontrivial subsystems of Hilbert spaces composed of both the control qubit and the target system at different times.  

\begin{figure}
\vspace{0.5 cm}
\begin{center}
\includegraphics[width=8cm]{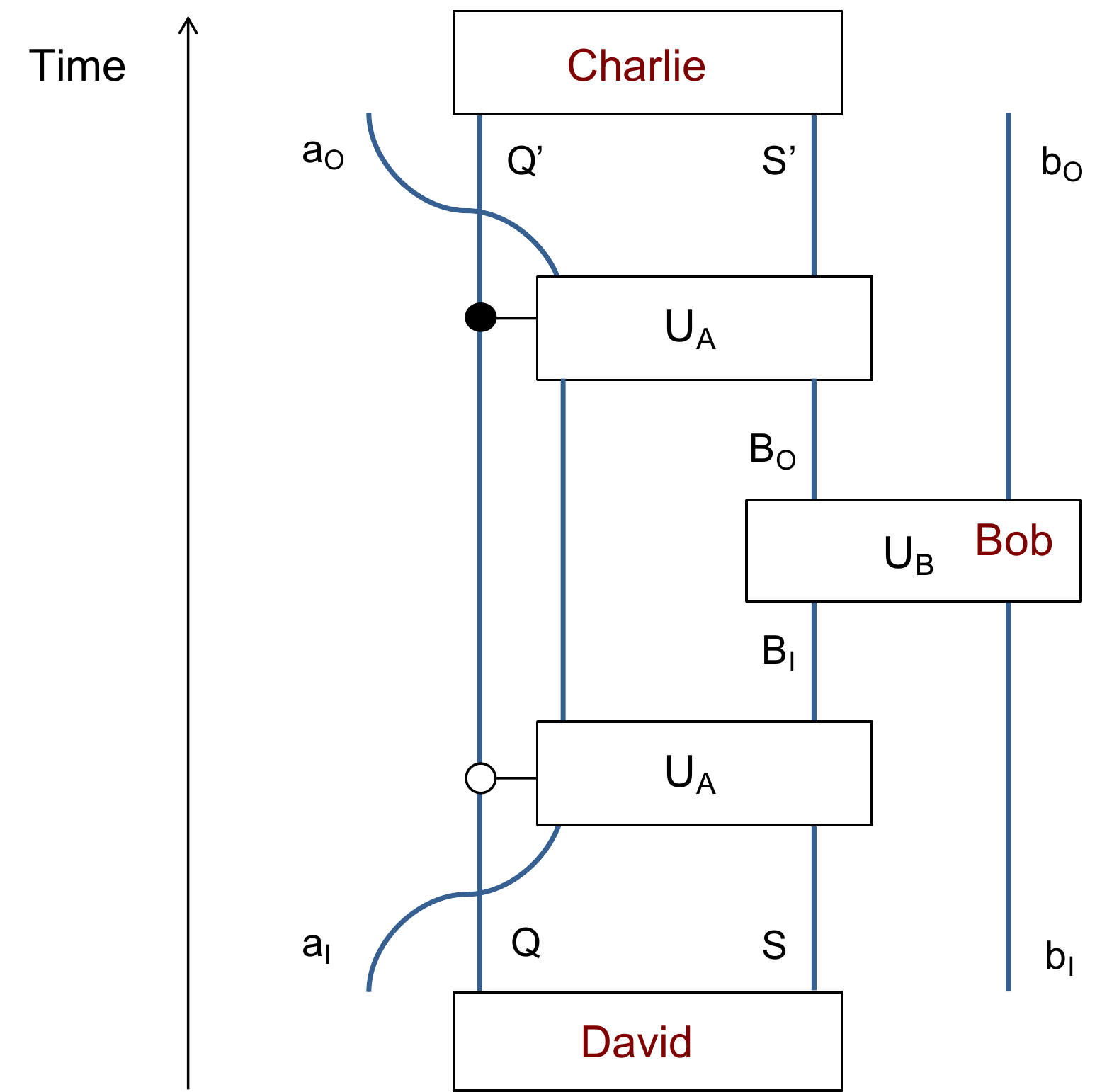}
\end{center}
\vspace{-0.5 cm}
\caption{\textbf{Temporal description of the implementation of the quantum SWITCH with Bob's operation at a fixed time.} From a temporal perspective, the experiment can be described by a sequence of unitaries acting on the control qubit and the target system plus the local auxiliary systems of Alice and Bob. Here, Bob's operation is implemented at a fixed time, while Alice's supposed operation is implemented by means of two controlled unitaries.} \label{fig:5}
\end{figure}

To simplify our analysis, we will restrict our attention to an implementation in which the operation of Bob is applied at a fixed time, while the operation of Alice may be applied before Bob's operation or after it, depending on the logical value of the control qubit. (The experiments reported in Refs.~\cite{Procopio, Rubino1, Rubino2} are based on a symmetric setup with respect to the possible times of the operations of Alice and Bob, and below we will comment on how our result looks in that case as well.) From a temporal perspective, the experiment has the circuit structure depicted in Fig. \ref{fig:5}, where the unitaries are given at the Hilbert-space level, the operations of David and Charlie are left unspecified, and we use a standard graphical notation for controlled unitaries \cite{NielsenChuang}, where a black dot represents conditioning on the state $|1\rangle\langle 1|$ and a white dot on the state $|0\rangle\langle 0|$. 

\textit{Remark.} The circuit in Fig. \ref{fig:5} is drawn with respect to the control qubit and the target system, which in the conceived realization are the path and internal degrees of freedom of a particle. Neither of these systems is a spatially localized system. One may ask how this description relates to a system decomposition that reflects the spatiotemporal configuration. Assuming that Alice and Bob reside at separate spatial locations and that either vacuum or one particle can enter their laboratory at any given time, a natural choice of a spatially local system to associate with each laboratory would be the $d+1$-dimensional Hilbert space that is a direct sum of the local vacuum and the $d$-dimensional Hilbert space of the internal degrees of freedom of the particle. The full Hilbert space of the joint system of Alice and Bob at a given time is then the tensor product of these two local Hilbert spaces. However, in the described experimental setup only a subspace of this Hilbert space is ever populated---the one corresponding to the presence of exactly one particle, which could be in either Alice's or Bob's laboratory or some superposition of the two locations. This subspace decomposes into a tensor product of a two-dimensional subsystem encoding the location of the particle (or the path degree of freedom) and the $d$-dimensional subsystem corresponding to the internal degrees of freedom. Thus, with respect to a spatially local choice of systems, the control qubit and the target system are factors of a proper subspace of the full Hilbert space. As noted in the introduction, this is the most general realization of a system within a given Hilbert space, and it will be essential in the generalization discussed in Sec. \ref{section7}. Note, however, that the present argument is independent of how precisely the circuit in Fig. \ref{fig:5} is realized---one may even think that the depicted systems are spatially local and that the realization consists of directly applying the controlled operations displayed on the figure by turning on and off suitable fields. 
\begin{figure}
\vspace{0.5 cm}
\begin{center}
\includegraphics[width=6cm]{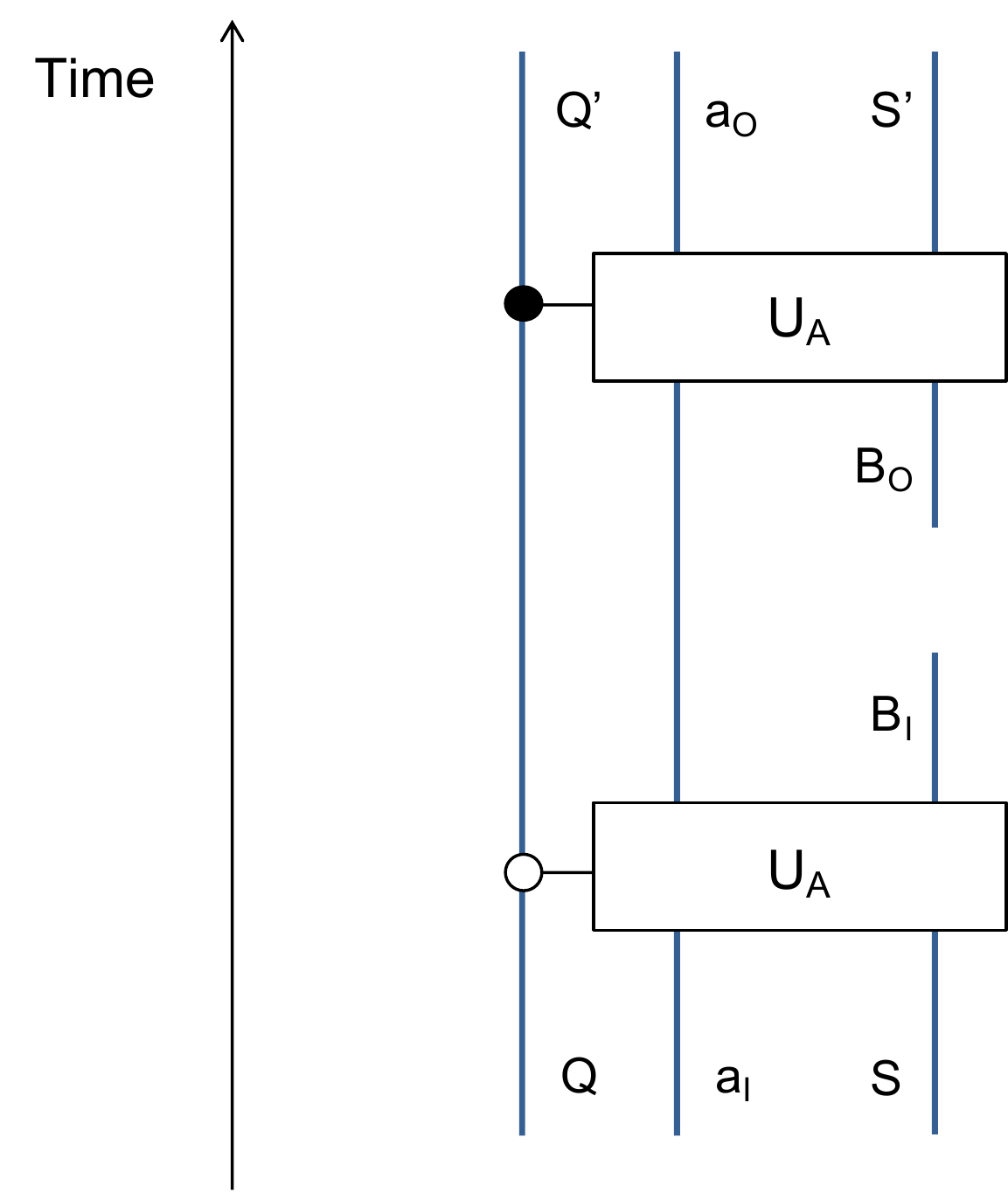}
\end{center}
\vspace{-0.5 cm}
\caption{\textbf{Circuit fragment containing Alice's operation.} Removing the operations of David, Bob, and Charlie from Fig. \ref{fig:5}, we are left with this circuit fragment in which Alice's operation is supposedly taking place in some sense.} \label{fig:6}
\end{figure}

\begin{figure}
\vspace{0.5 cm}
\begin{center}
\includegraphics[width=7cm]{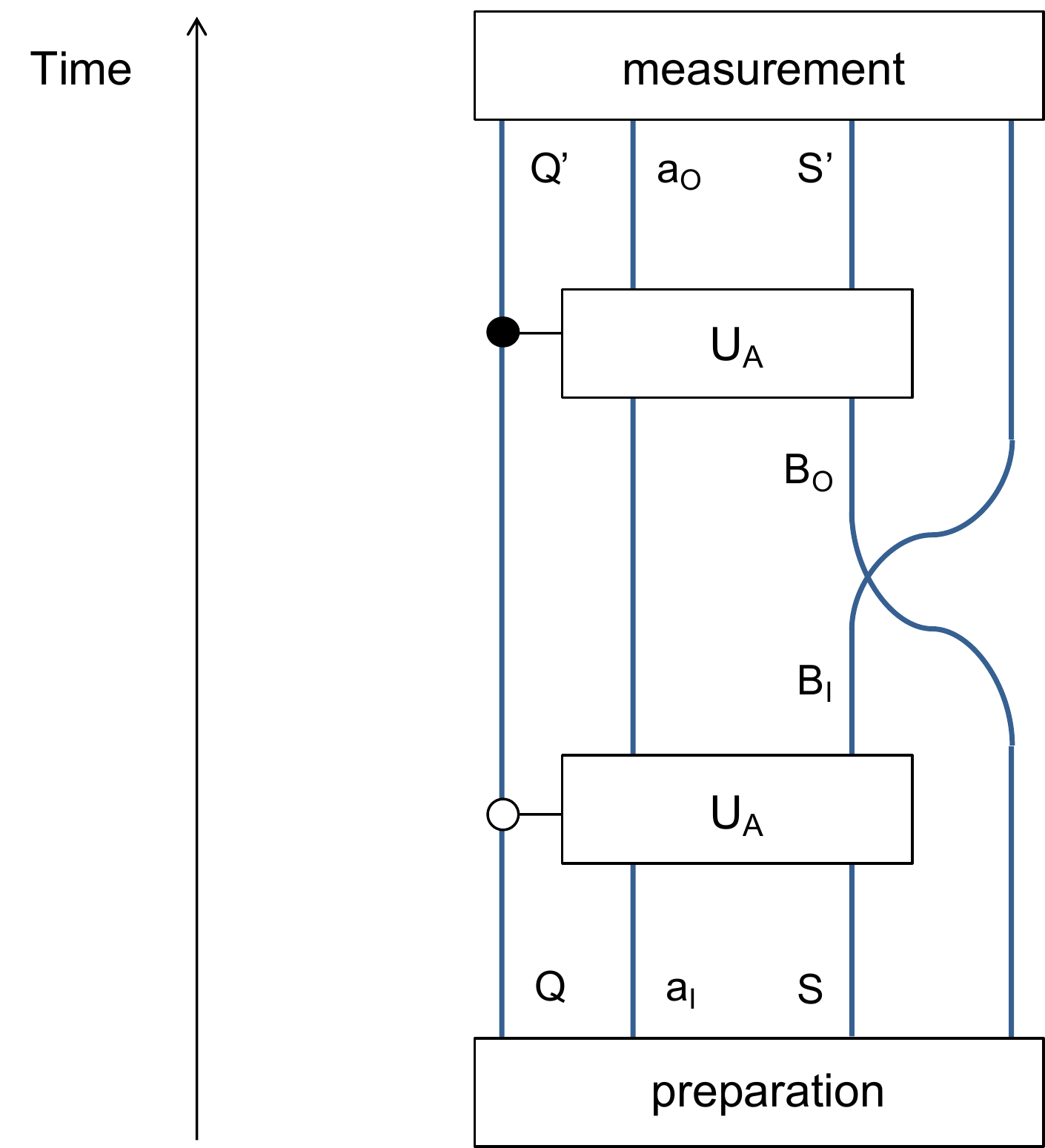}
\end{center}
\vspace{-0.5 cm}
\caption{\textbf{Tomography of circuit fragments and time-delocalized operations: an example.} The operation implemented by a circuit fragment can be tested via tomography similarly to any standard operation, by `pulling' all input wires to the past and all output wires to the future via identity transformations, as depicted here for the fragment from Fig. \ref{fig:6}. Via the same general scheme, one can also perform tomography of time-delocalized operations, such as $U_A^{a_{I}A_{I}\rightarrow a_{O}A_{O}}$ in Eq. \eqref{unitarystructure}. This simply requires preparing suitable states in the input subsystem $a_{I}A_{I}$, and applying suitable measurements on the output subsystem $a_{O}A_{O}$.} \label{fig:7}
\end{figure}

By assumption, the input and output systems of David, Bob, and Charlie are associated with fixed times. We know what it means operationally to say that David, Bob, and Charlie perform specific operations with these input and output systems: we can test this by feeding suitable states in their input systems and performing suitable measurements on their output systems. Alice's supposed operation $\mathcal{U}_A^{a_{I}A_{I}\rightarrow a_{O}A_{O}}$, on the other hand, is some still unidentified part of the circuit fragment in Fig. \ref{fig:6} (the description in Fig. \ref{fig:6} is given in terms of the transformations at the Hilbert-space level). It is clear that such a fragment is itself a quantum operation from the composite input system $a_{I}QSB_{O}$ to the composite output system $a_{O}Q'S'B_{I}$ (which is described by the theory of quantum combs \cite{networks}; in particular, it is a 1-comb). In this case, the operation is a unitary channel whose unitary matrix at the Hilbert-space level is 
\begin{gather}
U_{\textrm{SWITCH}}^{a_{I}QSB_{O}\rightarrow a_{O} Q'S'B_{I} }{(U_A)}=\notag \\
|0\rangle^{Q'} \langle 0|^{Q} \otimes  U_A^{a_{I}S \rightarrow a_{O}B_{I}} \otimes \id^{B_{O}\rightarrow S'} 
 \notag\\ + |1\rangle^{Q'} \langle 1|^{Q} \otimes U_A^{a_{I}B_{O}\rightarrow  a_{O} S'} \otimes \id^{S \rightarrow B_{I}},   \label{unitarycomb}
\end{gather}
which can be obtained by multiplying the unitary gates that make up the circuit fragment. This statement can also be verified through tomography by feeding suitable states in the joint input and performing suitable measurements on the joint output, as depicted in Fig. \ref{fig:7}.

Our first main observation is that the unitary in Eq.~\eqref{unitarycomb} can be written in the form   
\begin{gather}
U_{\textrm{SWITCH}}^{a_{I}QSB_{O}\rightarrow a_{O}Q'S'B_{I}}{(U_A)} =U_A^{a_{I}A_{I}\rightarrow a_{O}A_{O}}\otimes \id^{\overline{A_{I}}\rightarrow \overline{A_{O}}},\label{unitarystructure}
\end{gather} 
or, equivalently, the unitary channel of the quantum comb can be written in the form 
\begin{gather}
\mathcal{U}_{\textrm{SWITCH}}^{a_{I}QSB_{O}\rightarrow a_{O}Q'S'B_{I}}{(U_A)} =\mathcal{U}_A^{a_{I}A_{I}\rightarrow a_{O}A_{O}}\otimes \mathcal{I}^{\overline{A_{I}}\rightarrow \overline{A_{O}}},\label{unitarystructureChannel}
\end{gather} 
where $\mathcal{I}$ denotes the identity channel. Here, $\mathcal{H}^{A_{I}}$ is a tensor factor of $\mathcal{H}^{QSB_{O}}$ that has dimension $d$ and $\mathcal{H}^{\overline{A_{I}}}$ is its cofactor ($\mathcal{H}^{QSB_{O}} = \mathcal{H}^{A_{I}}\otimes \mathcal{H}^{\overline{A_{I}}}$), and similarly, $\mathcal{H}^{A_{O}}$ is a tensor factor of $\mathcal{H}^{Q'S'B_{I}}$ that has the dimension of $d$ and $\mathcal{H}^{\overline{A_{O}}}$ is its co-factor ($\mathcal{H}^{Q'S'B_{I}} = \mathcal{H}^{A_{O}}\otimes \mathcal{H}^{\overline{A_{O}}}$). The factor $\mathcal{H}^{A_{I}}$ is defined by the algebra of operators of the form 
\begin{gather}
O^{A_{I}} \equiv |0\rangle\langle 0|^{Q}\otimes O^{S}\otimes \id^{B_{O}}+ |1\rangle\langle 1|^{Q}\otimes \id^{S}\otimes O^{B_{O}},\label{Ain}
\end{gather}
and the factor $\mathcal{H}^{A_{O}}$ by the algebra of operators of the form 
 \begin{gather}
O^{A_{O}} \equiv |0\rangle\langle 0|^{Q'}\otimes \id^{S'} \otimes O^{B_{I}}+ |1\rangle\langle 1|^{Q'}\otimes O^{S'} \otimes \id^{B_{I}}. \label{Aout}
\end{gather}
These are the input and output systems of Alice's operation.  

To see this, notice that \cite{Esteban}, for any $U_A$, 
\begin{gather}
U_{\textrm{SWITCH}}^{a_{I}QSB_{O} \rightarrow a_{O}Q'S'B_{I}} {(U_A)}\notag\\
= \textrm{C-SWAP}^{Q'S'B_{I}} (U_A^{a_{I}S\rightarrow a_{O}B_{I}}\otimes \id^{QB_{O}\rightarrow Q'S'}) \textrm{C-SWAP}^{QSB_{O}} ,\label{swappedU}
\end{gather} 
where $\textrm{C-SWAP}^{XYZ}$ denotes the controlled-SWAP unitary operator (which is also Hermitian) with control qubit $X$ and target systems $Y$ and $Z$ \cite{SWAP}. This can be verified from the expression for $U_{\textrm{SWITCH}}^{a_{I}QSB_{O}\rightarrow a_{O} Q'S'B_{I} }(U_A )$ in Eq.~\eqref{unitarycomb}. The operator sandwiched between the two C-SWAP operators on the right-hand side of Eq.~\eqref{swappedU} has the same form as the right-hand side of Eq.~\eqref{unitarystructure}, except that $A_{I}$ is replaced by $S$, and $A_{O}$ is replaced by $B_{I}$. Since the first C-SWAP maps some subsystem of $QSB_{O}$ (call it $A_{I}$) to the subsystem $S$, and the second C-SWAP maps the subsystem $B_{O}$ to some subsystem of $Q'S'B_{I}$ (call it $A_{O}$), the action of $U_{\textrm{SWITCH}}^{a_{I}QSB_{O}\rightarrow a_{O} Q'S'B_{I} }(U_A )$ is precisely the one given in Eq.~\eqref{unitarystructure} with respect to $A_{I}$ and $A_{O}$. The subsystems $A_{I}$ and $A_{O}$ are related to $S$ and $B_1$ via the above C-SWAP transformations, which is equivalent to the relations \eqref{Ain} and \eqref{Aout} for the operators on these subsystems.  

Having identified the subsystems $A_{I}$ and $A_{O}$, the operation performed from $A_{I}$ to $A_{O}$ becomes a directly testable fact---we can verify it by preparing suitable states on $A_{I}$ and performing suitable measurements on $A_{O}$, which fits within the general scheme depicted in Fig. \ref{fig:7}. 

\begin{figure}
	\vspace{0.5 cm}
	\begin{center}
		\includegraphics[width=6.5cm]{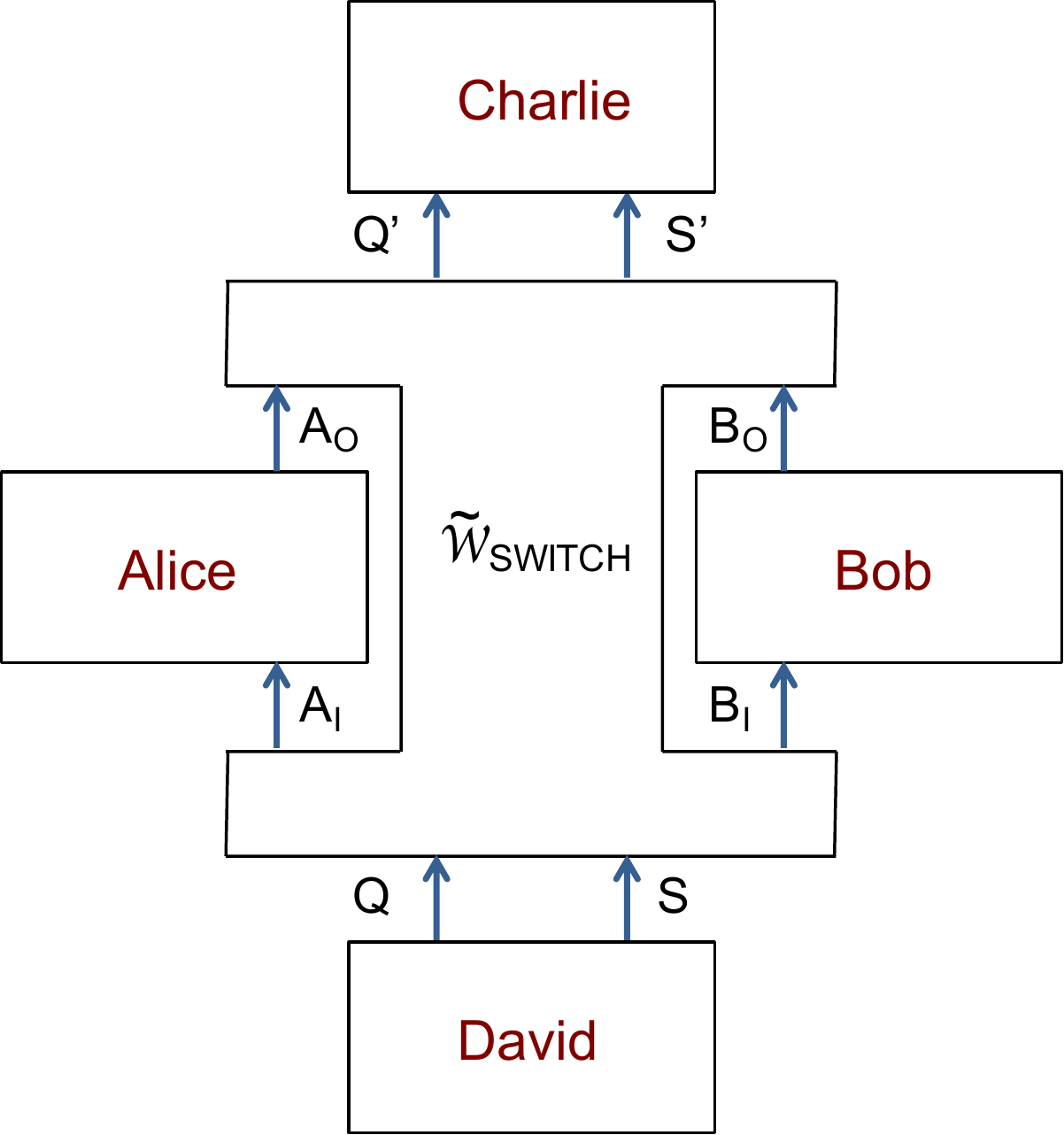}
	\end{center}
	\vspace{-0.5 cm}
	\caption{\textbf{The quantum SWITCH as an irreducible cyclic circuit.} With respect to the time-delocalized subsystems $A_{I}$ and $A_{O}$, the discussed implementation has the structure of a circuit with a cycle. This is trivially true for any experiment involving a set of first-order operations since any process matrix is equivalent to a channel, as discussed earlier. However, when the local operations are part of a standard acyclic circuit, the box of the process can be further decomposed into a finer-grained circuit fragment such that the loop is explicitly broken. But when the process is causally nonseparable, as is the case here, the circuit cannot be decomposed into a finer-grained acyclic circuit or a probabilistic mixture of acyclic circuits.} \label{fig:8}
\end{figure}

Note that the subsystem $A_{I}$ has a nontrivial `spread' over $B_{O}$ in the sense that it is not a subsystem of the complement $QS$. Similarly, $A_{O}$ has a nontrivial spread over $B_{I}$. But $B_{I}$ is in the causal past of $B_{O}$. In spite of this, it is possible to perform an arbitrary standard operation from $A_{I}$ to $A_{O}$. While this fact may appear counterintuitive, it is important to stress that, at least in this case, it is not specific to quantum mechanics---it holds irrespectively of whether we initialize $Q$ in a quantum superposition or a classical probabilistic mixture of the logical states, and even has an analogue in the case where all systems are classical. In those classical cases, however, it is possible to think that the logical observable on $Q$ has a definite value and hence the `true' input system of Alice's operation is not a fixed one but either $S$ or $B_{O}$, depending on the control bit (formally speaking, this is equivalent to associating $A_{I}$ with a tensor factor of two different subspaces of $QSB_{O}$), and similarly the output system is either $B_{I}$ or $S'$. However, when $Q$ is initialized in a quantum superposition and Charlie is allowed to perform arbitrary operations on $Q'S'$, such an interpretation is not possible anymore. In general, with respect to the input and output systems $A_{I}$ and $A_{O}$ that we have identified, the experiment has the cyclic circuit structure depicted in Fig. \ref{fig:8}, where the channel connected to the operations of the four parties cannot be decomposed into a mixture of channels where the cycle might break. (Note that for simplicity we have defined $A_I$ as directly connected to $QSB_O$ without any temporal separation, and $A_O$ as directly connected to $B_IQ'S'$, i.e., the central box in Fig. \ref{fig:8} is not strictly `bulky' from a temporal perspective, despite what is suggested by the graphical representation. However, we could also have realizations where the central box is `bulky' if we insert identity channels with nonzero time spans at the places of the systems $Q$, $S$, $B_I$, $B_O$, $Q'$, $S'$ in Fig. \ref{fig:5}.) As first pointed out by Chiribella \textit{et al.} \cite{Chiribella12}, such a cyclic structure is a necessary property of the quantum SWITCH as a higher-order transformation on the operations of Alice and Bob \cite{no-go}. By identifying precisely these operations, our analysis makes the cyclic structure explicit, vindicating the interpretation of the described implementation as a realization of the quantum SWITCH. (For more on this point, see Sec. \ref{section4}.)

\begin{figure}
	\vspace{0.5 cm}
	\begin{center}
		\includegraphics[width=8cm]{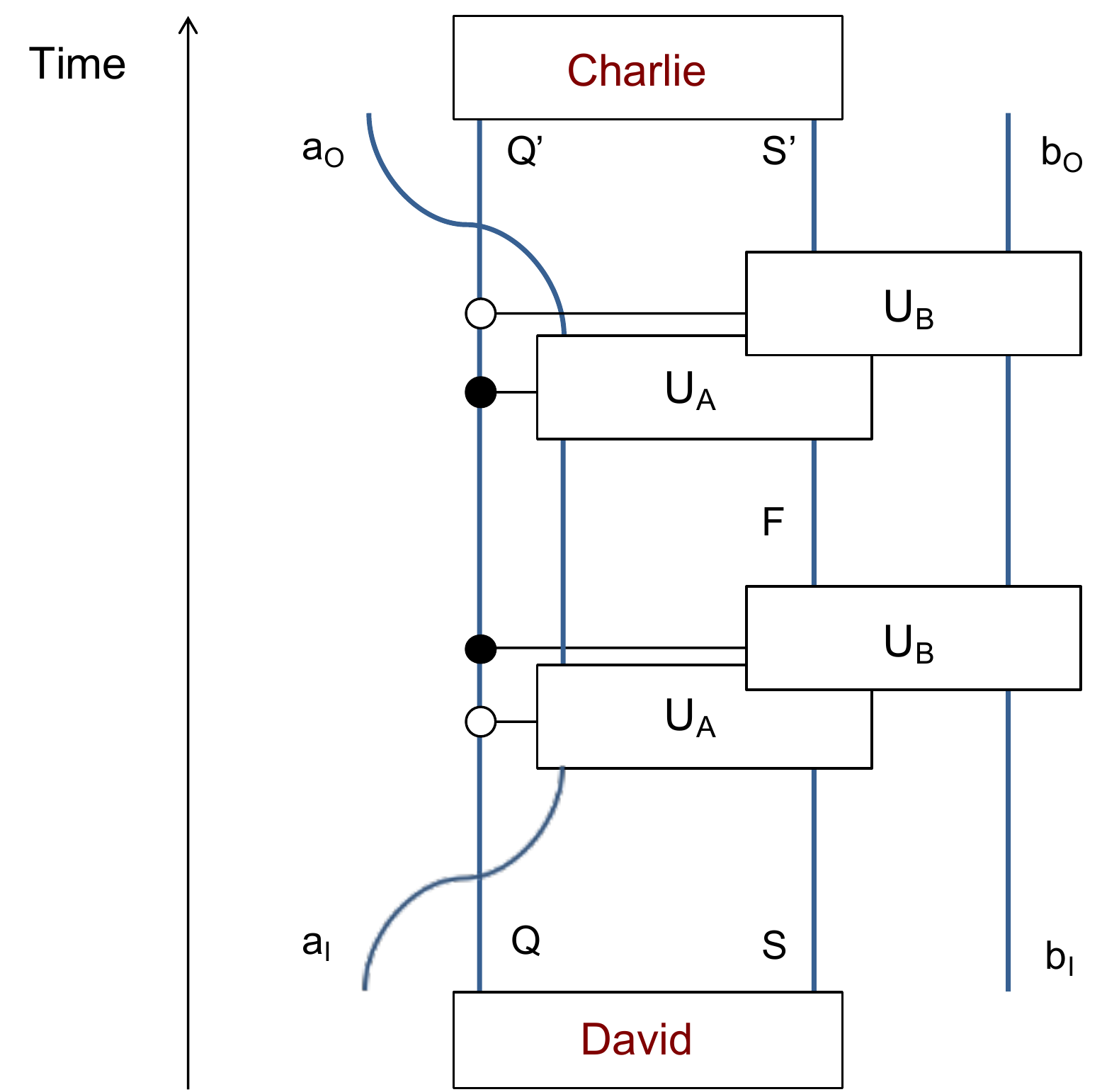}
	\end{center}
	\vspace{-0.5 cm}
	\caption{\textbf{Temporal description of a symmetric implementations of the quantum SWITCH.} The experimental implementations of the quantum SWITCH in Refs. \cite{Procopio, Rubino1, Rubino2} are symmetric with respect to the possible times of the operations of Alice and Bob. In this case, the temporal description of the experiment in terms of the control qubit and the target system has the form displayed here.} \label{fig:9}
\end{figure}

Let us now comment on the input and output systems of Alice and Bob in the case of a symmetric implementation, such as the one in the experiments \cite{Procopio, Rubino1, Rubino2}. Again, identifying the control and target systems, we may draw the overall circuit as in Fig. \ref{fig:9}, where the controlled unitaries are such that either Alice's or Bob's operation takes place at a given time, depending on the logical value of the control qubit (the controlled unitaries are displayed with a slight shift rather than simultaneous for graphical clarity). Following analogous analysis to the one presented earlier, one sees that Alice's input and output systems are given by 

\begin{gather}
O^{A_{I}} \equiv |0\rangle\langle 0|^{Q}\otimes O^{S}\otimes \id^{F} + |1\rangle\langle 1|^{Q}\otimes \id^{S}\otimes O^{F},\label{AinSymmetric}
\end{gather}
\begin{gather}
O^{A_{O}} \equiv |0\rangle\langle 0|^{Q'}\otimes O^{F}\otimes \id^{S'}  + |1\rangle\langle 1|^{Q'}\otimes \id^{F}\otimes O^{S'}, \label{Aoutsymmetric}
\end{gather}
and Bob's input and output systems are given by 
\begin{gather}
O^{B_{I}} \equiv |0\rangle\langle 0|^{Q}\otimes \id^{S}\otimes O^{F} + |1\rangle\langle 1|^{Q}\otimes O^{S}\otimes \id^{F},\label{Bin}
\end{gather}
\begin{gather}
O^{B_{O}} \equiv |0\rangle\langle 0|^{Q'}\otimes \id^{F}\otimes O^{S'}  + |1\rangle\langle 1|^{Q'}\otimes O^{F}\otimes \id^{S'}. \label{Bout}
\end{gather}

\section{On the notion of `realization', and a comment on resources}\label{section4}

It is instructive to discuss in more detail the sense in which we regard the previous results as justifying the interpretation of those experiments as `realizations' of the higher-order process. The perspective taken in this paper is that a higher-order process, such as the quantum SWITCH, is a mathematical concept defined in terms of CP maps over Hilbert spaces. To say that such a mathematical concept has a physical realization means that there can be an experiment in which every element of the description of the concept has a physical counterpart that itself can be considered a realization of that element. What we have shown is that, in the discussed types of experiments, every element of the description of the process -- Hilbert spaces, and operations on them -- has a physical counterpart. What may still be debated is whether we are willing to regard that counterpart as a `realization' of the element.

When it comes to experiments of the kind discussed in the previous section, their interpretation as realizations of standard circuits composed of time-localized operations is well established. However, if one proposes an alternative interpretation of the same experiment based on a new notion of realization of a CP map, it is natural to question whether we are willing to accept this new notion. But in this case, the proposed notion is not really new. A time-delocalized CP map is a trivial combination of two already accepted notions: (1) quantum combs as CP maps on Hilbert spaces that are tensor products of Hilbert spaces associated with definite times, and (2) subsystems as realizations of Hilbert spaces. We therefore regard the proposed interpretation in full agreement with and equally sound as the standard one. 

Of course, one may be interested in realizations of quantum operations that obey further conditions. For instance, one may ask that a quantum operation is realized by the action of a spatially local physical box on the internal degrees of freedom of a particle passing through it. However, from our perspective this is just a particular kind of realization, and there is no reason to assume that all realizations of a quantum operation must be of this kind. For example, operations acting on the Hilbert spaces of fields, which include the vacuum state, are just as meaningful. The main goal of the present work is to point out that, in agreement with the established theory, we can have more general types of systems and operations than those usually considered, which offers a more general way in which we can contemplate realizations of processes. 

A common objection to the assertion that an implementation of the kind described in the previous section can be regarded as a realization of the quantum SWITCH concerns the fact that the temporal circuit description of the experiment involves multiple controlled versions of the input operations (or, more generally, operations on a larger Hilbert space that includes the vacuum). Indeed, one of the celebrated differences between the quantum SWITCH and higher-order processes in which the input operations are compatible with definite causal order is that the quantum SWITCH produces a specific operation that is a function of $U_A$ and $U_B$ with a single use of each $U_A$ and $U_B$, whereas processes with a definite causal order require at least two uses of $U_A$ or $U_B$ to produce the same operation. An example of an implementation that uses two copies of one input operation and a single copy of the other one was described in Eq. (18) of \cite{Chiribella12}. Such implementations are commonly understood as `simulations' of the quantum SWITCH, as opposed to realizations. Although we have seen that with respect to the relevant systems the implementations in the previous section are exactly equivalent to the quantum SWITCH applied on single copies of $U_A$ and $U_B$, one may still be concerned by the fact that the temporal circuit contains two controlled-$U_A$ operations, where a single use of controlled-$U_A$ is a more powerful resource than a single use of $U_A$ (because we can obtain the latter from the former by preparing a suitable state on the control system). Are we not using too powerful resources for this implementation to be considered a realization?

To understand this issue, it is crucial to clarify the context in which implementations with multiple copies are regarded as simulations \cite{Chiribella12}. The context is that we are given access to time-localized versions of two unknown unitaries, $f$ and $g$ (we use the notation corresponding to Fig. (18) of \cite{Chiribella12}), and we want to produce the unitary $U_{\textrm{SWITCH}} (f,g)$. Using standard circuits, we would need to call at least one of the unknown unitaries, say $g$, twice. However, if we could apply the higher-order operation \textit{quantum SWITCH} on such unitaries, we would achieve the result with a single use of each $f$ and $g$. Note that since by assumption these are time-localized versions of $f$ and $g$, the ability to apply a quantum SWITCH on them would imply the ability to send information back in time (this can be inferred from the steps in the proof of Proposition 1 in Ref. \cite{Chiribella12}). The sense in which an implementation such as the one in Fig. (18) of \cite{Chiribella12} can be regarded as a simulation is that it simulates the effect of having access to a quantum SWITCH of this kind plus a single use of each $f$ and $g$, for the purpose of producing $U_{\textrm{SWITCH}} (f,g)$.

In the present paper, we are not concerned with the task of producing $U_{\textrm{SWITCH}} (f,g)$ by using time-localized versions of $f$ and $g$. We are asking whether the higher-order transformation \textit{quantum SWITCH} can be said to have a physical realization. We have seen that the answer is positive, but the `slots' in which the input operations must be `plugged' are given by time-delocalized subsystems. Such a quantum SWITCH cannot be applied on time-localized operations (and in particular does not allow signaling back in time). It is nevertheless a realization of the higher-order transformation in the sense we have described. Furthermore, it can be regarded as a resource in comparison to causally separable processes in an appropriately redefined context. For example, instead of being given access to time-localized versions of the input unitaries $f$ and $g$, we could be given access to these operations applied on time-delocalized subsystems (or systems of our choice, as long as this is physically permissible). In such a scenario, the process connecting the operations can still be causally separable. Indeed, one can see that we can create any causally separable process between Alice, Bob, Charlie, and David, when Alice's operation is time-delocalized as in the scheme in Fig. \ref{fig:5}, by inserting in that circuit suitable operations after David, before and after Bob, and before Charlie. (Essentially, this is because we can fix the order between Alice and Bob by fixing the state of the control qubit right before the quantum 1-comb that contains the operation of Alice.) Therefore, the comparison between the quantum SWITCH and causally separable processes is meaningful in this scenario, and the proven advantages hold. 

It is clear that a time-delocalized version of a given operation is a different physical resource than a time-localized version. One could ask how these resources relate to each other, e.g., what resources of one kind are needed to obtain a resource of the other kind. (A rigorous formulation of this question would require defining the resources to which we have access for free, and the way we can use all resources.) For instance, we have seen that having access to two uses of time-localized controlled-$g$, where $g$ is unknown, allows us to create one time-delocalized copy of $g$ of the kind on which a quantum SWITCH can be applied. Obviously, the former resource is strictly more powerful than the latter since we cannot use a single time-delocalized versions of $g$ to obtain two copies of controlled-$g$ that we could use in a circuit independently. In practice, we could be given access to either of these two resources. There is a common tendency to think that the gates that make up the temporal circuit in implementations of the quantum SWITCH (e.g., Fig. \ref{fig:5}) represent `the' resource of interest. This is likely because these gates are associated with single uses of physical devices (e.g., optical elements) that in the laboratory context could naturally be applied independently. But the latter is not mandatory---we can imagine that we only have access to time-delocalized versions of $g$ rather than time-localized versions of controlled-$g$ if, for example, the operations of interest are provided by another party. Regarding the intuition that operations are associated with single uses of physical devices, this idea is compatible with time-delocalized operations too: consider a machine, which upon pressing of a button applies one Hamiltonian pulse, followed, after some time, by another Hamiltonian pulse, so as to implement the sequence of gates in Fig. \ref{fig:6} \cite{endnoteX}. Such a device could itself be used multiple times in principle, and one could ask how many times it has been used. In this concrete example, it is used exactly once by construction. 

Finally, it is instructive to look at what the time-delocalized subsystem perspective can tell us about simulations of the quantum SWITCH (in the sense described earlier) via multiple copies of some input operation, such as the one in Eq. (18) of Ref. \cite{Chiribella12}. For that example, it is easy to see  that there are suitable time-delocalized subsystems with respect to which the circuit takes the form of a quantum SWITCH on the two input operations $f$ and $g$, plus a second copy of $g$ applied in parallel on some additional system. One can also see that it is physically possible to implement only the quantum SWITCH part without the residual copy of $g$, which from a temporal perspective would look like applying a controlled-$g$ twice, rather than $g$ twice in the circuit in Eq. (18) of Ref. \cite{Chiribella12}, similarly to Fig. \ref{fig:6}. In other words, the implementation with two copies of $g$ contains a realization of the quantum SWITCH on time-delocalized subsystems, plus a residual time-delocalized version of $g$. In particular, we see that two time-localized copies of $g$ can be used to produce a time-delocalized copy of $g$ of the type required for the quantum SWITCH, plus a residue.

\section{General processes based on coherent control of the times of the operations}\label{section5}

\begin{figure}
	\vspace{0.5 cm}
	\begin{center}
		\includegraphics[width=6.5cm]{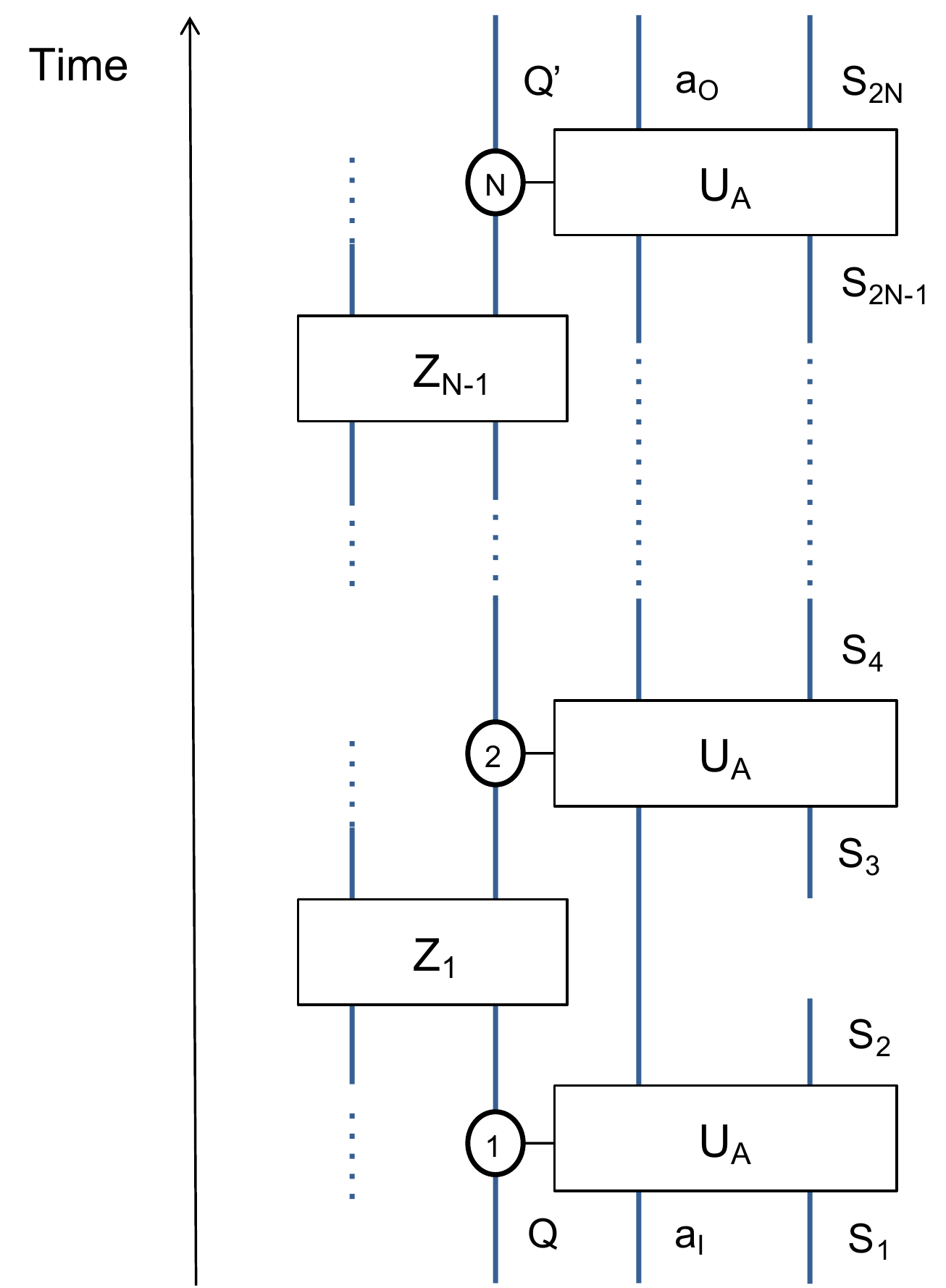}
	\end{center}
	\vspace{-0.5 cm}
	\caption{\textbf{A time-delocalized operation realized through a sequence of controlled gates.} Here, a controlled unitary depicted with ``$i$'' in the circle on the control system represents a controlled unitary that is `triggered' if and only if the logical value of the control observable is $i$. We can permit arbitrary gates $Z_j$ to act on the control system at any time (here shown in between the controlled gates for clarity) as long as they commute with the control observable. We can also permit implementations in which the gates that make up the circuit fragment are modified from the ones in this figure in a way that preserves the  transformation realized by the fragment. A fragment of this type can be seen to give rise to a unitary that contains the factor $U_A^{a_{I}A_{I}\rightarrow a_{O}A_{O}}$, where $A_{I}$ and $A_{O}$ are defined in Eqs.\eqref{AinGEN} and \eqref{AoutGEN}.} \label{fig:10}
\end{figure}

The result of Sec. \ref{section3} generalizes straightforwardly to multipartite process implementations where the operation of each party may be delocalized over a number of possible times by means of controlled unitaries conditioned on a logical observable (most generally defined by a complete set of orthogonal projectors $\{P_{i}\}_{i=1}^N$ over some control system) that is preserved during a period containing these times, as described in Fig. \ref{fig:10}. [Note that we can permit gates $Z_i$ to act on the control system during the period of the controlled gates, as shown in the figure, as long as they commute with the logical observable. We can further modify the gates that make up the circuit fragment in any way that leaves the transformation realized by the fragment invariant (e.g., appending local unitaries on the control system to the controlled operations, which are undone by their inverses in the past or in the future).] The input system of such an operation would be a subsystem of ${QS_1S_3\cdots S_{2N-1}}$ defined by the algebra of operators 
\begin{gather}
O^{A_{I}} \equiv \sum_{i=1}^{N}P_{i}^{Q}\otimes O^{S_{2i-1}}\otimes \id^{\overline{S_{2i-1}}},\label{AinGEN}
\end{gather}
where $\overline{S_{2i-1}}$ denotes the complementary subsystem of $S_{2i-1}$ in $S_1S_3\cdots S_{2N-1}$,
and the output system would be a subsystem of ${Q'S_2S_4\cdots S_{2N}}$ defined by the algebra of operators 
 \begin{gather}
O^{A_{O}} \equiv \sum_{i=1}^{N}P_i^{Q'}\otimes O^{S_{2i}}\otimes  \id^{\overline{S_{2i} } }, \label{AoutGEN}
\end{gather}
where $\overline{S_{2i} } $ denotes the complementary subsystem of $S_{2i}$ in $S_2S_4\cdots S_{2N}$.

In Refs. \cite{OG, AraujoWitness} it was shown that a class of multipartite processes containing operations realized through this procedure cannot violate causal inequalities. It seems likely that no process based on such operations would be able to violate causal inequalities, for reasons similar to those used in the arguments in Refs. \cite{OG, AraujoWitness}. This conjecture is left open for future investigation.

\section{Unitarily extendible bipartite processes}\label{section6}

\begin{figure}
	\vspace{0.5 cm}
	\begin{center}
		\includegraphics[width=11cm]{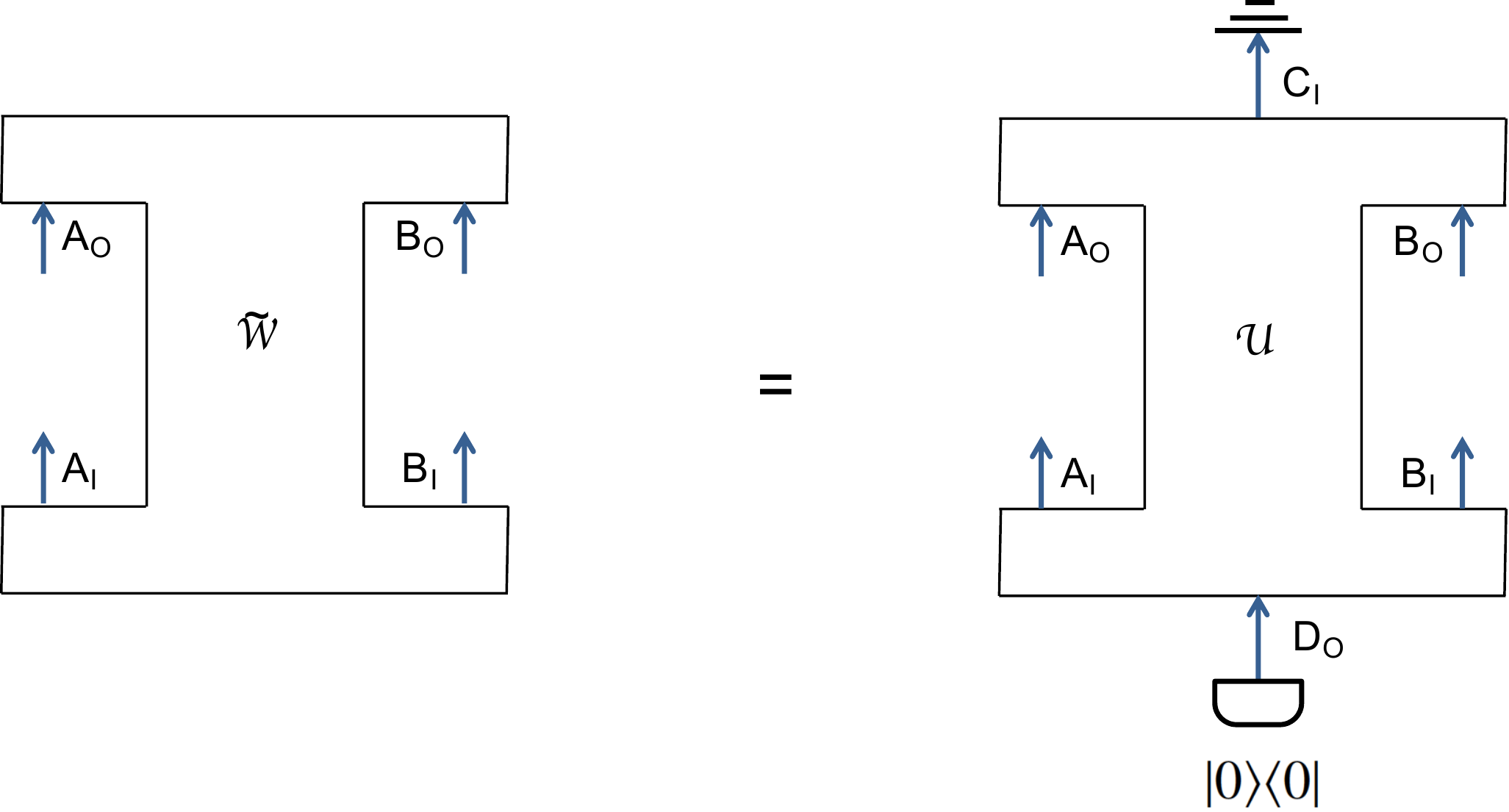}
	\end{center}
	\vspace{-0.5 cm}
	\caption{\textbf{Unitary extendibility of a bipartite process.} A bipartite process obeys the unitary extension postulate proposed in Ref. \cite{AraujoPostulate} if and only if there exists a four-partite process of the kind on the right-hand side of this figure that is a unitary channel, such that the original process is obtained from this four-partite process conditionally on David preparing his output system in some pure state $|0\rangle\langle 0|$, and Charlie tracing out his input system. The unitary extendibility of multipartite processes is defined analogously \cite{AraujoPostulate}.} \label{fig:11}
\end{figure}

\begin{figure}
	\vspace{0.5 cm}
	\begin{center}
		\includegraphics[width=7cm]{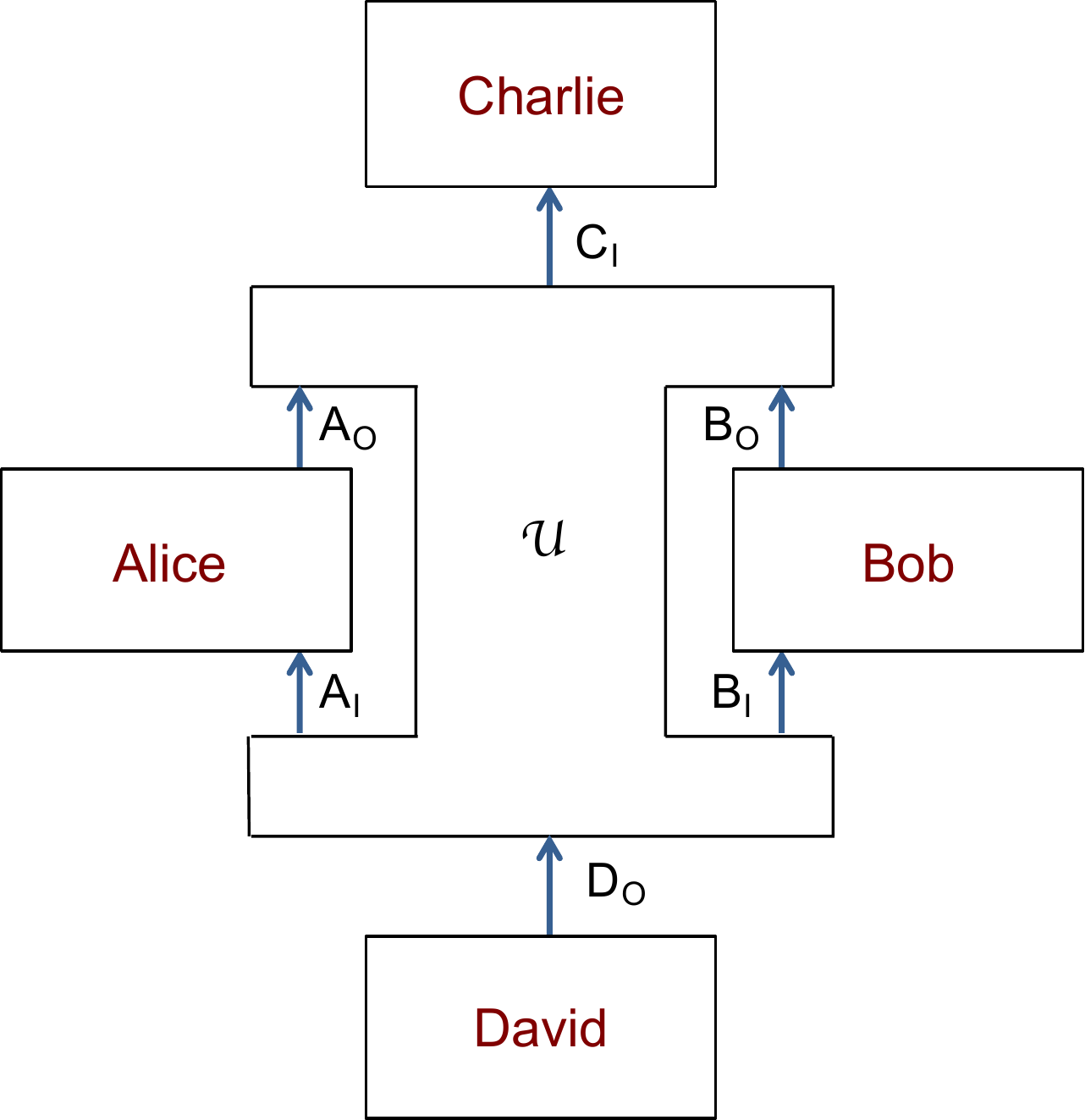}
	\end{center}
	\vspace{-0.5 cm}
	\caption{\textbf{A class of unitary processes that have a realization on time-delocalized subsystems.} We consider the class of four-partite processes of the kind depicted in this figure, which are equivalent to unitary channels from $D_OA_OB_O$ to $A_IB_IC_I$. We show that they admit realizations on time-delocalized subsystems (see Theorem 1).} \label{fig:12}
\end{figure}

In Ref. \cite{AraujoPostulate}, Ara\'{u}jo \textit{et al.} proposed a postulate that might distinguish processes that have a physical realization from those that do not, showing that certain processes violating causal inequalities fail to respect it. The postulate says that any physically admissible process should be possible to obtain from an extended process involving two additional parties, one having only a nontrivial output system and the other one only a nontrivial input system, in a way analogous to the one in Fig. \ref{fig:11} for the case of two parties, where the extended process is equivalent to a unitary channel from $D_OA_OB_O$ to $A_IB_IC_I$. (Note that it is crucial in this definition that the unitary channel $\mathcal{U}$ is not just any unitary dilation of the channel $\widetilde{\mathcal{W}}$, but is itself a valid process, i.e., the transpose of its CJ matrix is a valid process matrix.) As we have seen in the previous section, the quantum SWITCH is a four-partite process of the kind on the right-hand side of Fig. \ref{fig:11}, and hence it provides a unitary extension for a class of bipartite processes (obtained for different choices of the state on ${D_O}$). These bipartite processes, however, are causally separable \cite{OG, AraujoWitness}, even though the quantum SWITCH is not. At the time of the initial submission of this paper, it was not know whether there are causally nonseparable bipartite processes that obey the above postulate \cite{CyclicProof}. But by the same argument sketched earlier for the case of the quantum SWITCH, the unitary extension of any bipartite process would in general be causally nonseparable except in the extreme case where the bipartite process is compatible with a fixed causal order (i.e., it is nonsignaling from Alice to Bob or from Bob to Alice). Here, we show that all bipartite processes that obey the postulate, together with their unitary extensions, have a realization on suitable time-delocalized quantum subsystems.\\

\begin{figure}
	\vspace{0.5 cm}
	\begin{center}
		\includegraphics[width=14cm]{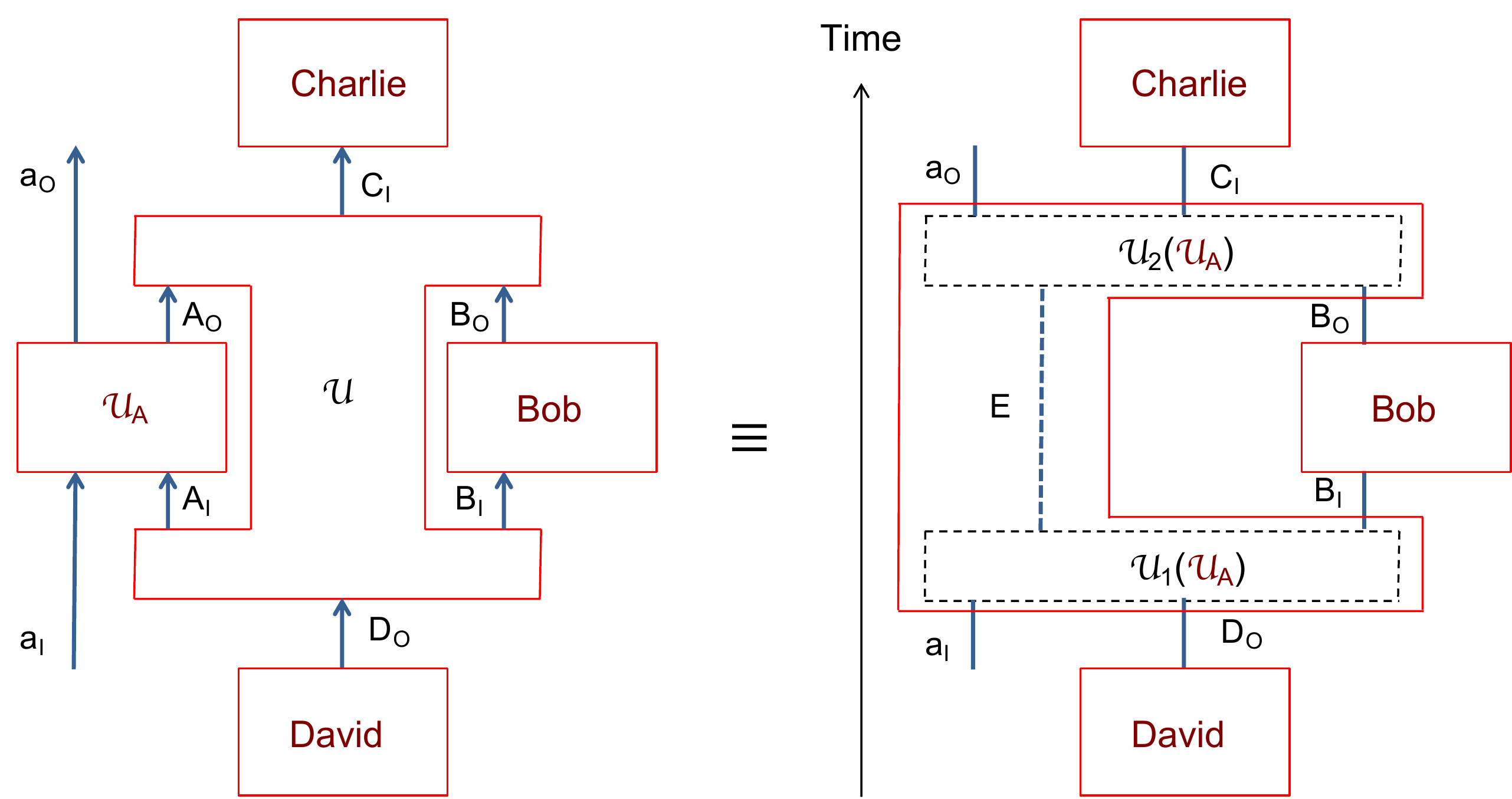}
	\end{center}
	\vspace{-0.5 cm}
	\caption{\textbf{Equivalence of specific circuits and unitary higher-order processes.} According to Theorem 1, there is a subsystem $A_I$ of the time-local systems $D_OB_O$ depicted on the right-hand side, and a subsystem $A_O$ of the time-local systems $B_IC_I$ depicted on the right-hand side, such that for appropriately chosen unitary channels $\mathcal{U}_1(\mathcal{U}_A)$ and $\mathcal{U}_2(\mathcal{U}_A)$, the circuit on the right-hand side is equivalent to the process on the left-hand side (both understood verifiable box-wise).} \label{fig:13}
\end{figure}

\textbf{Theorem 1 (Realizability of unitary extensions of bipartite processes).} Consider a unitary four-partite process of the kind depicted in Fig. \ref{fig:12}. In agreement with the known quantum mechanics, there exist time-local systems $D_O$, $B_I$, $B_O$, $C_I$, ordered in time in this order, a subsystem $A_I$ of $D_OB_O$, and a subsystem $A_O$ of $B_IC_I$, such that the circuit-like process depicted on the left-hand side of Fig. \ref{fig:13}, for arbitrary operations of David, Bob, and Charlie (left unspecified in the figure for simplicity), and an arbitrary unitary operation $\mathcal{U}_A^{a_IA_I \rightarrow a_OA_O }$ of Alice (which can be used to obtain arbitrary operations in Alice's slot), where $a_I$ and $a_O$ are ancillary input and output systems, admits an experimental implementation on these systems without post-selection. The implementation has a temporal description as on the right-hand side of Fig. \ref{fig:13}, where the unitaries $\mathcal{U}_1$ and $\mathcal{U}_2$ are functions of $\mathcal{U}_A$.\\

\textit{Remark on graphical notation.} In Fig. \ref{fig:13}, as well as in all of the following figures in this and the next section, the boxes with (red) solid boundaries and the solid wires connecting them denote the objects about which we are stating equivalences. The dashed fragments inside the solid boxes describe compositions of operations that can yield the respective total transformations in the solid boxes.

\begin{figure}
	\vspace{0.5 cm}
	\begin{center}
		\includegraphics[width=12cm]{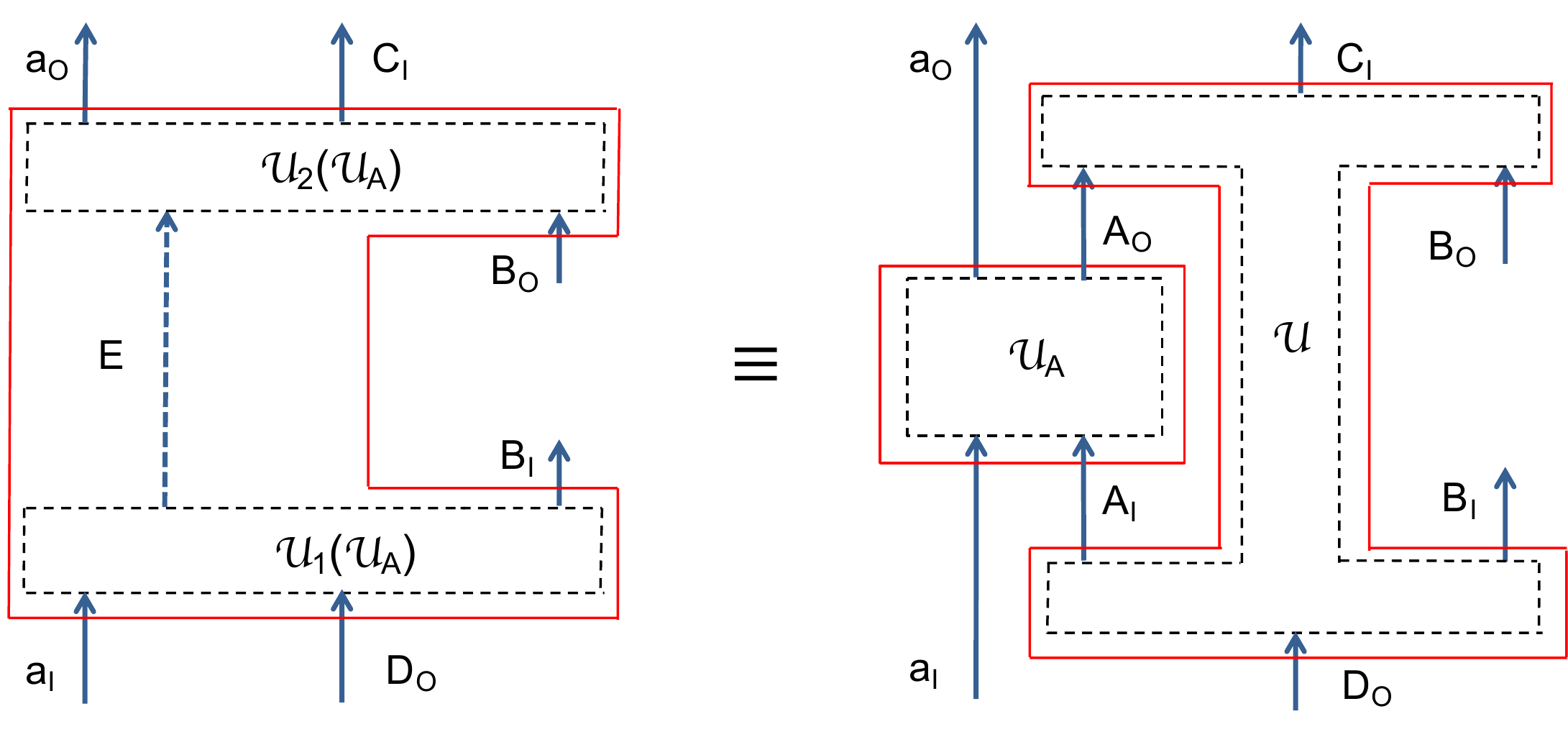}
	\end{center}
	\vspace{-0.5 cm}
	\caption{\textbf{The main equivalence behind Theorem 1.} The proof of Theorem 1 is based on proving the equivalence between the process fragment on the right-hand side and the quantum 1-comb on the left-hand side, for suitable subsystems $A_I$ and $A_O$ of the boundary systems on the left-hand side.} \label{fig:14}
\end{figure}

\textbf{Proof}. In agreement with the known quantum mechanics, one can find time-local systems $D_O$, $B_I$, $B_O$, $C_I$, ordered in time as on the right-hand side of Fig. \ref{fig:13}, such that it is possible to implement arbitrary quantum operations between these systems (the time intervals between the systems may need to be sufficiently large so as to allow universal transformations to be realized in agreement with the Hamiltonian that governs the fundamental fields). This ensures the realizability of the operations of David, Bob, and Charlie, as depicted on the right-hand side of Fig. \ref{fig:13}, as well as of arbitrary operations between them. Therefore, the theorem would follow if we can show the existence of a subsystem $A_I$ of $D_OB_O$ and a subsystem $A_O$ of $B_IC_I$, and unitary transformations $\mathcal{U}_1(\mathcal{U}_A)$ and  $\mathcal{U}_1(\mathcal{U}_A)$ dependent on $\mathcal{U}_A$, such that the equivalence in Fig. \ref{fig:14} holds for every $\mathcal{U}_A$.

\begin{figure}
	\vspace{0.5 cm}
	\begin{center}
		\includegraphics[width=13.5cm]{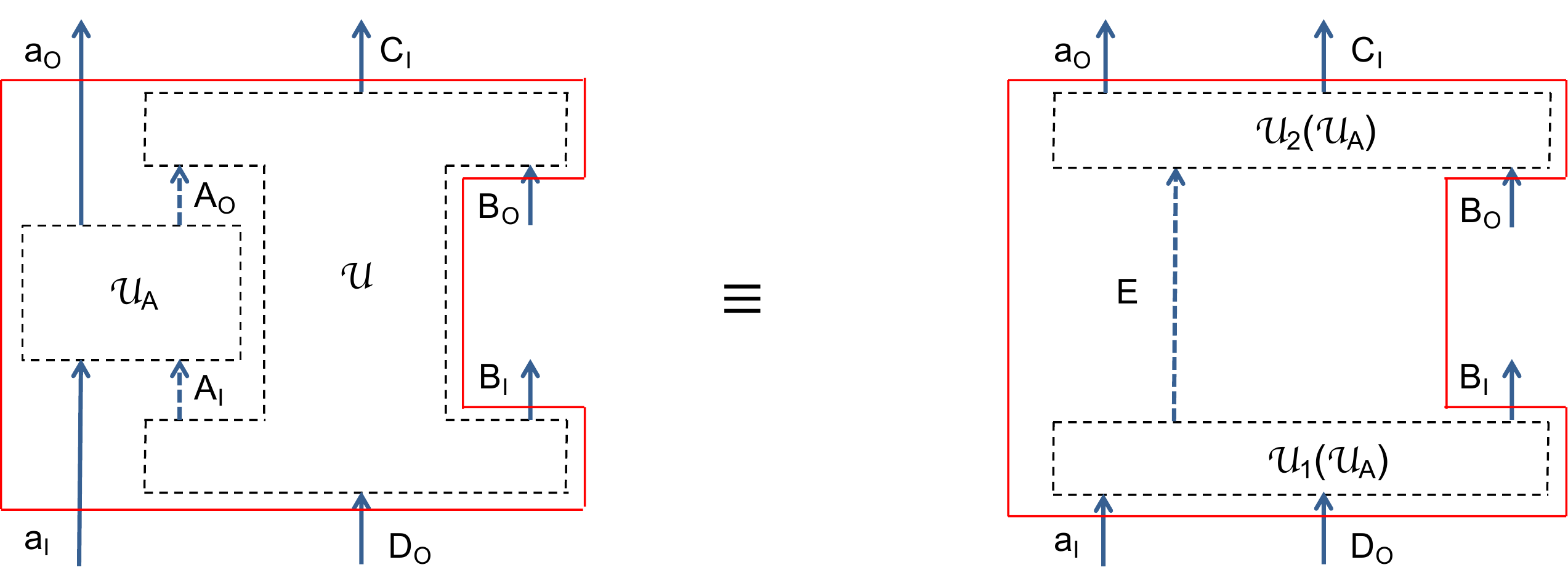}
	\end{center}
	\vspace{-0.5 cm}
	\caption{\textbf{The composition of Alice's unitary channel with the unitary process yields a unitary quantum 1-comb.} For any choice of extended unitary channel on Alice's side, the process reduces to a supermap from Bob's operation to a new operation, which can be realized physically by a unitary circuit fragment connected to Bob's operation \cite{Chiribella08}. This is a special case of Fig. \ref{fig:16}.} \label{fig:15}
\end{figure}

\begin{figure}
	\vspace{0.5 cm}
	\begin{center}
		\includegraphics[width=13.5cm]{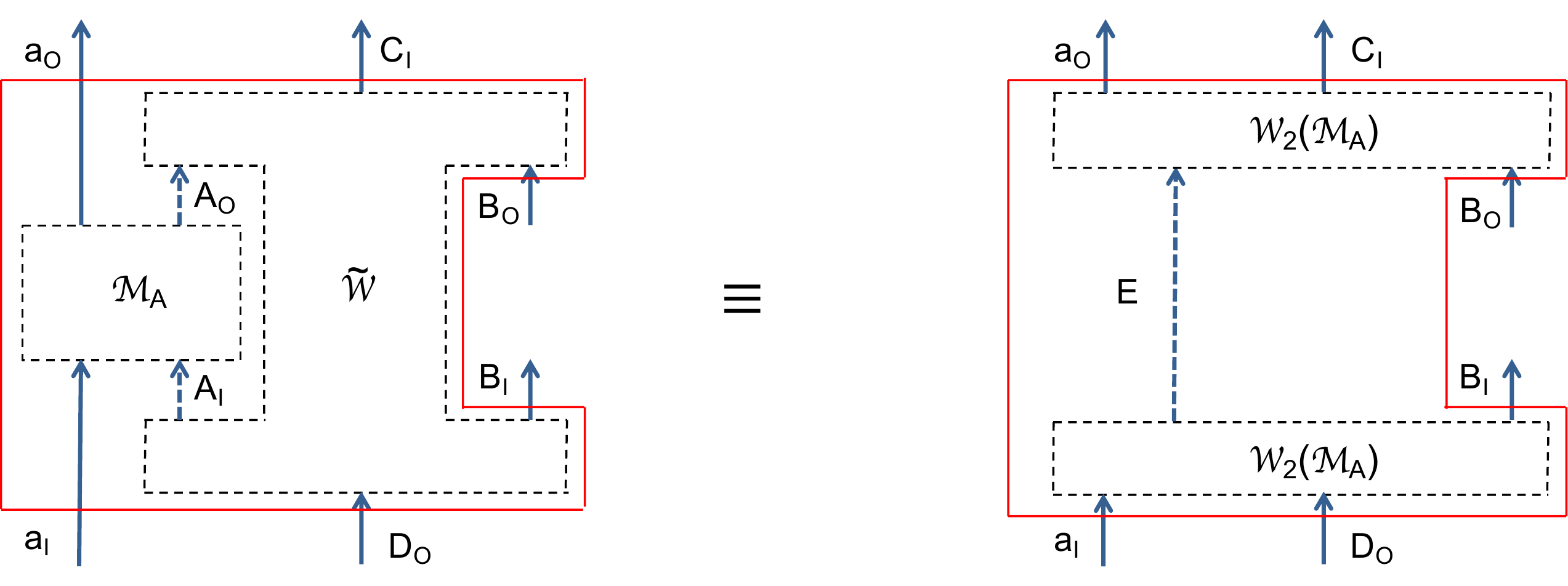}
	\end{center}
	\vspace{-0.5 cm}
	\caption{\textbf{The composition of Alice's channel with a process of the considered kind yields a quantum 1-comb.} For any choice of operation on Alice's side, a process of the type considered here reduces to a supermap from Bob's operation to a new operation, which can be realized physically by a circuit fragment connected to Bob's operation \cite{Chiribella08}.} \label{fig:16}
\end{figure}

Note that if the equivalence in Fig. \ref{fig:14} holds, then it must in particular hold that the resultant transformation from $a_ID_OB_O$ to $a_OB_IC_I$ on the left-hand side is equal to the corresponding resultant transformation on the right-hand side, as depicted in Fig. \ref{fig:15}. But for any process of the type considered here (not necessarily unitary), and for any choice of channel for Alice (not necessarily unitary), the transformation resulting from the composition of Alice's channel with the process is equivalent to a supermap \cite{Chiribella08} on Bob's operation. This can be easily seen from the fact that a process, by definition, is normalized on all deterministic operations performed by the parties. As shown in Ref. \cite{networks}, any deterministic supermap on a single operation is equivalent to a quantum 1-comb and can be realized by a circuit fragment composed of isometries, where some subsystem is discarded at the end. This means that in general we have the equivalence in Fig. \ref{fig:16} for suitable channels $\mathcal{W}_1(\mathcal{M}_A)$ and $\mathcal{W}_2(\mathcal{M}_A)$ that are functions of $\mathcal{M}_A$. In the case when both $\widetilde{\mathcal{W}}$ and $\mathcal{M}_A$ are unitary (and hence their composition is unitary \cite{AraujoPostulate}), these channels can themselves be taken unitary. Indeed, as shown in Ref. \cite{networks}, the construction of the circuit fragment that implements a quantum comb can be taken such that it achieves the minimal Stinespring dilation \cite{Stinespring} for the full comb, which in the case of a unitary comb yields a circuit fragment composed of unitary channels as in Fig. \ref{fig:15}. Moreover, these unitary channels are unique up to a joint transformation that is formally equivalent to a change of basis for the system $E$. (For how to construct such a circuit fragment, see Ref.~\cite{networks}.) Therefore, if there exists a circuit fragment as on the right-hand side of Fig. \ref{fig:14} such that the equivalence in Fig. \ref{fig:14} holds, that circuit fragment must belong to the equivalence class arising from the construction in Ref.~\cite{networks}. We will show that there exist subsystems $A_I$ of $D_OB_O$ and $A_O$ of $B_IC_I$, such that for any unitary fragment obtained through this construction, the equivalence in Fig. \ref{fig:14} holds, i.e., any such fragment provides a realization of the process fragment on the left-hand side of Fig. \ref{fig:14}.

\begin{figure}
	\vspace{0.5 cm}
	\begin{center}
		\includegraphics[width=13cm]{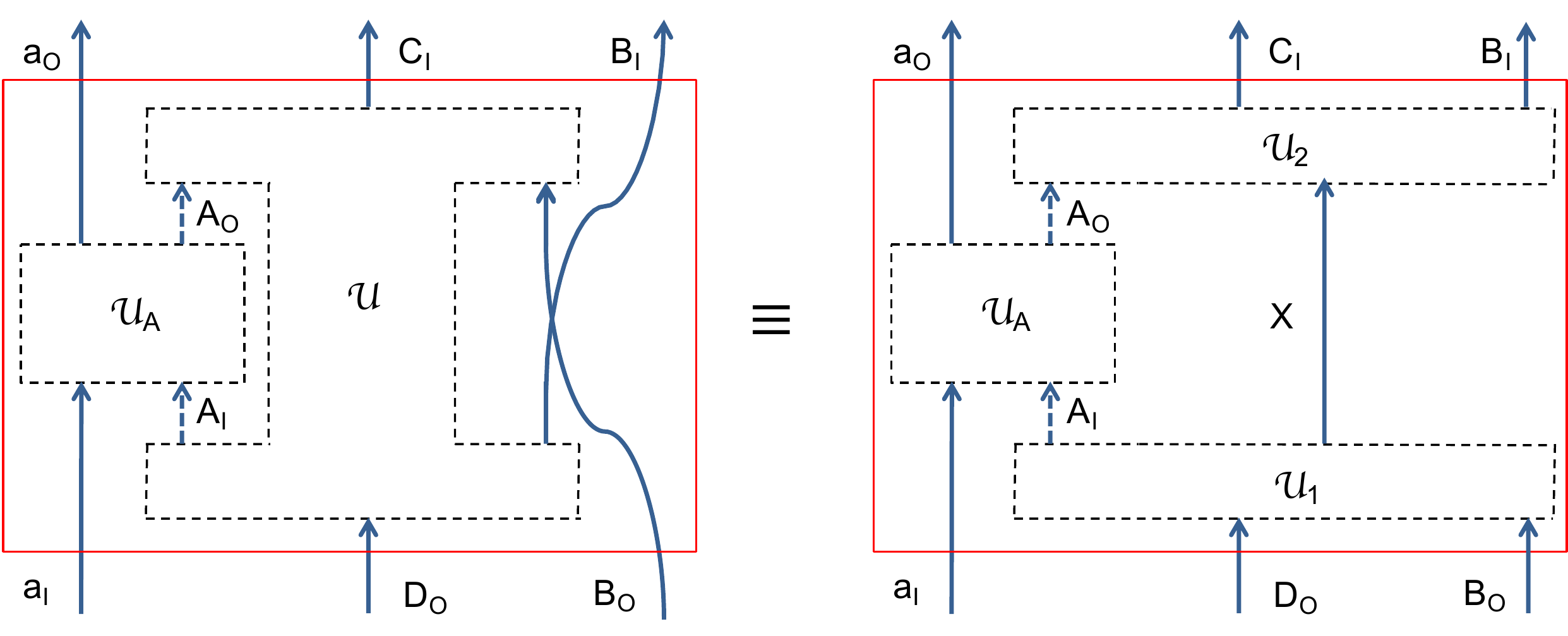}
	\end{center}
	\vspace{-0.5 cm}
	\caption{\textbf{Alternative realization of the unitary channel of the quantum comb in Fig.\ref{fig:15}.} Deforming the left-hand side of Fig. \ref{fig:15} by `pulling' up and down the `wires' corresponding to Bob's input and output systems is formally equivalent to plugging the SWAP operation in Bob's slot, which must yield the result of a supermap applied on Alice's operation, implying the alternative realization inside the red box on the right-hand side of this figure.} \label{fig:17a}
\end{figure}

The proof that such subsystems exist uses crucially the fact that the channel of a process does not allow signaling from the output of one party to the input of the same party, which means that the unitary channel $\mathcal{U}^{D_OA_OB_O\rightarrow A_IB_IC_I}$ maps $A_{O}$ isomorphically \cite{isomorphic} to some subsystem $\tilde{A}_{O}$ of $B_{I}C_{I}$, and similarly maps some subsystem $\tilde{A}_{I}$ of $D_{O}B_{O}$ isomorphically to $A_{I}$. To formally identify these subsystems, it is convenient to deform the fragment contained in the red box on the left-hand side of Fig. \ref{fig:15} by `pulling' the input and output `wires' of Bob up and down, respectively, so as to put the red box in a more standard input-to-output layout, as on the left-hand side of Fig.~\ref{fig:17a}. The overall transformation in the red box does not change under this graphical deformation. But we may also interpret the deformation as describing the plugging of a SWAP transformation into Bob's slot in the process fragment inside the red box, which by the same argument as above must yield a supermap on Alice's operation that has a realization as a unitary circuit fragment. In other words, we have the equivalence displayed in Fig. \ref{fig:17a} for specific unitary channels $\mathcal{U}_1$ and $\mathcal{U}_2$, which depend on the process $U$, and which, similarly to above, are uniquely defined up to a joint transformation formally equivalent to a change of basis on the intermediary system $X$ (see below). Looking at the circuit inside the red box on the right-hand side of Fig. \ref{fig:17a}, we see that there must be a subsystem $\tilde{A}_{I}$ of $D_{O}B_{O}$ which is mapped isomorphically to $A_{I}$ by the channel $\mathcal{U}_1$. It is defined by the algebra of operators of the form 
\begin{gather}
O^{\tilde{A}_{I}}\equiv U_1^{\dagger}(O^{A_{I}} \otimes \id^X )U_1,\label{iso1}
\end{gather}
where $U_1$ is the unitary associated with the channel $\mathcal{U}_1$ at the Hilbert-space level. 
Similarly, the system $A_{O}$ is mapped isomorphically to a subsystem $\tilde{A}_{O}$ of $B_{I}C_{I}$, which is defined by the algebra of operators of the form
\begin{gather}
O^{\tilde{A}_{O}}\equiv U_2(O^{A_{O}} \otimes \id^X) U_2^{\dagger},\label{iso2}
\end{gather}
where $U_2$ is the unitary associated with the channel $\mathcal{U}_2$ at the Hilbert-space level. 
As noted already, the unitaries $U_1$ and $U_2$ are not unique: the quantum comb that they create when composed is invariant under transformations of the form $U_1\rightarrow U^XU_1$, $U_2\rightarrow U_2(U^X)^{\dagger}$, where $U^X$ is an arbitrary unitary on $X$. However, the operators  $O^{\tilde{A}_{I}}$ and $O^{\tilde{A}_{O}}$ defined above are also invariant under such transformations and hence can be obtained starting from any valid choice of $U_1$ and $U_2$, i.e., the subsystems $\tilde{A}_{I}$ and $\tilde{A}_{O}$ are uniquely defined.

\begin{figure}
	\vspace{0.5 cm}
	\begin{center}
		\includegraphics[width=13cm]{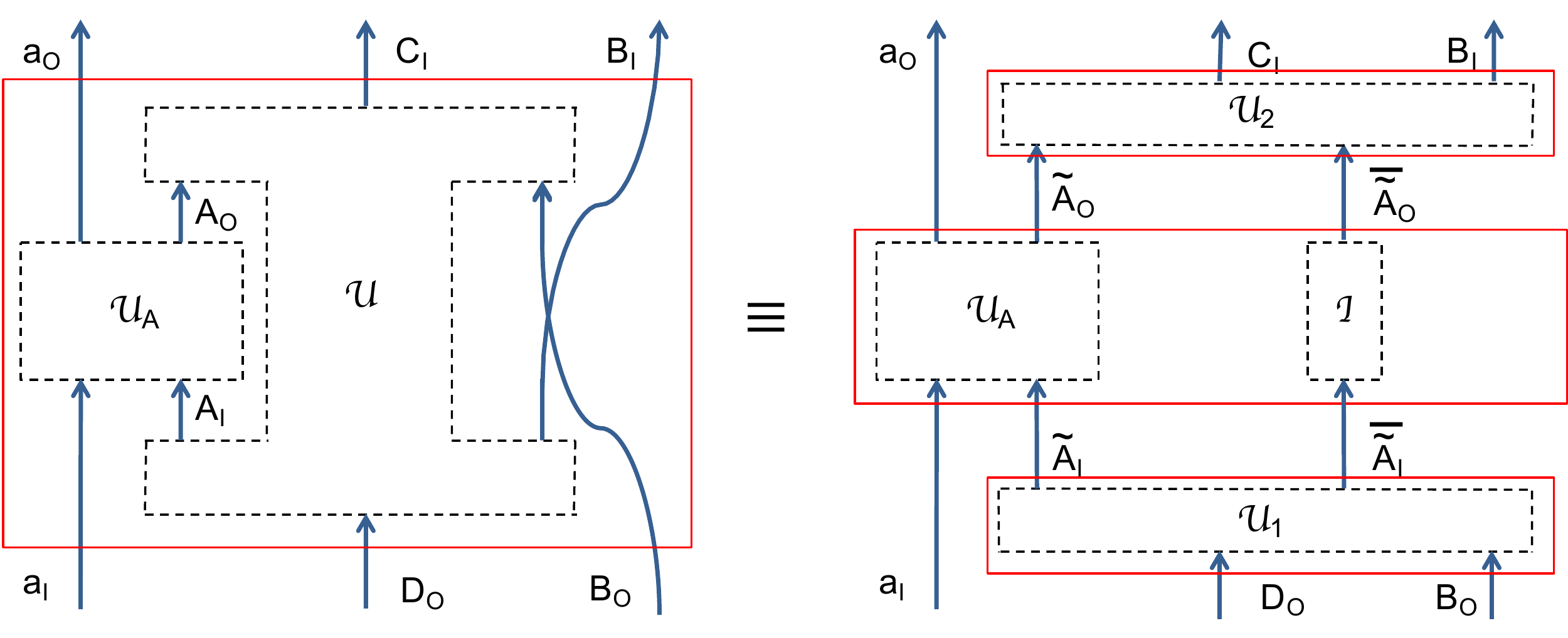}
	\end{center}
	\vspace{-0.5 cm}
	\caption{\textbf{Equivalent representation of the unitary channel in Fig. \ref{fig:17a}.} Using that the channel $\mathcal{U}_1$ in Fig. \ref{fig:17a} maps isomorphically some subsystem $\tilde{A}_I$ of $D_OB_O$ ($\mathcal{H}^{D_OB_O} = \mathcal{H}^{\tilde{A}_I}\otimes \mathcal{H}^{\overline{\tilde{A}}_I}$ ) onto $A_I$, and that the channel $\mathcal{U}_2$ in Fig. \ref{fig:17a} maps isomorphically the subsystem $A_O$ onto a subsystem $\tilde{A}_O$ of $B_IC_I$ ($\mathcal{H}^{B_IC_I} = \mathcal{H}^{\tilde{A}_O}\otimes \mathcal{H}^{\overline{\tilde{A}}_O}$), we conclude the equivalent representation depicted here, where each of the channels $\mathcal{U}_1$ and $\mathcal{U}_2$ on the right-hand side describes the link between two different factorizations of the same Hilbert space.} \label{fig:17b}
\end{figure}

Let $\overline{\tilde{A}}_I$ denote the co-subsystem of ${\tilde{A}}_I$ in $D_OB_O$ ($\mathcal{H}^{D_{O}B_{O}}= \mathcal{H}^{\tilde{A}_{I}} \otimes \mathcal{H}^{  \overline{\tilde{A}_{I}}}$), and let $\overline{\tilde{A}}_O$ denote the co-subsystem of ${\tilde{A}}_O$ in $B_IC_I$ ($\mathcal{H}^{B_{I}C_{I}}= \mathcal{H}^{\tilde{A}_{O}} \otimes \mathcal{H}^{  \overline{\tilde{A}_{O}}}$).
It is obvious from Fig. \ref{fig:17a} and the definition of $\tilde{A}_{I}$ and $\tilde{A}_{O}$ that the circuit fragment inside the box on the right-hand side transforms $\overline{\tilde{A}_{I}}$ to $X$ via a unitary, and then $X$ to $\overline{\tilde{A}_{I}}$ via another unitary, i.e., it maps unitarily $\overline{\tilde{A}_{I}}$ to $\overline{\tilde{A}_{I}}$. We can always choose an isomorphism between the initial and final subsystems (i.e., identify a basis for one with a basis for the other) such that the unitary connecting them is the identity. The unitary of the four-partite process matrix can then be written 
\begin{gather}
U^{D_{O}A_{O}B_{O}\rightarrow A_{I}B_{I}C_{I}} = \id^{\tilde{A}_{I} \rightarrow A_{I} }  \otimes \id^{A_{O} \rightarrow \tilde{A}_{O}}  \otimes \id^{\overline{\tilde{A}_{I}} \rightarrow \overline{\tilde{A}_{O}} }.
\end{gather}
This means that the unitary inside the box on the right-hand side of Fig. \ref{fig:17a} has the form 
\begin{gather}
U^{a_{I} D_{O}B_{O}    \rightarrow a_{O}  B_{I}C_{I}  }{(U_A)}= U_A^{a_{I}\tilde{A}_{I}\rightarrow  a_{O}\tilde{A}_{O}}\otimes \id^{\overline{\tilde{A}_{I}} \rightarrow \overline{\tilde{A}_{O}} }.\label{combfactorization}
\end{gather}

\begin{figure}
	\vspace{0.5 cm}
	\begin{center}
		\includegraphics[width=13cm]{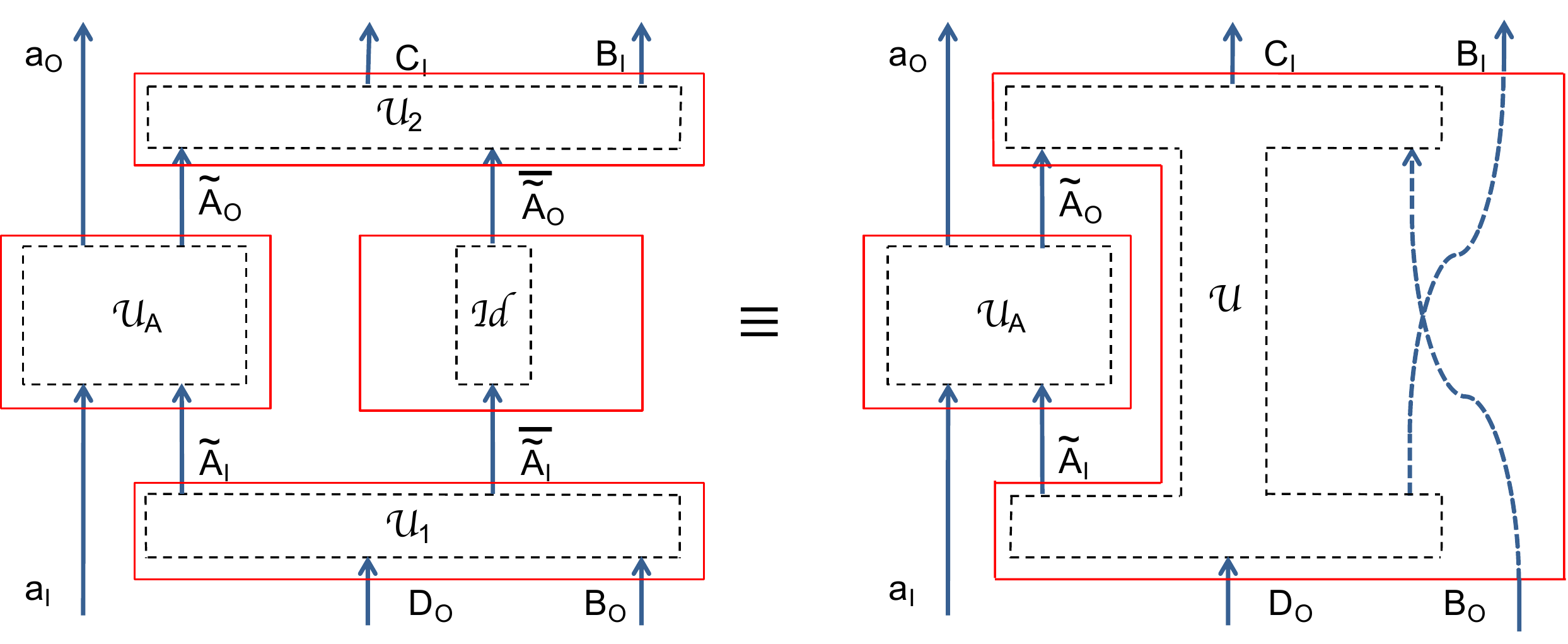}
	\end{center}
	\vspace{-0.5 cm}
	\caption{\textbf{Extracting Alice's operation.} Using that the middle box on the right-hand side of Fig. \ref{fig:17b} equals the tensor product of two operations, we may split it into two parallel boxes as on the left-hand here. Since the equivalence in Fig. \ref{fig:17b} must hold for all choices of Alice's unitary as and all operations that could be applied in the slots of the rest of the parties, whose correlations completely characterize the process connecting the slots of all parties, we conclude the equivalence displayed here.} \label{fig:18}
\end{figure}

We can therefore rewrite the transformation on the right-hand side of Fig. \ref{fig:17a} as on the right-hand side of Fig. \ref{fig:17b}, where the unitary channel $\mathcal{U}_1$ from $D_OB_O$ to $\tilde{A}_I\overline{\tilde{A}}_I$ is nothing but a trivial transformation (the identity) on the same system, which describes the relation between one subsystem decomposition of that system and another, and similarly for $\mathcal{U}_2$. Such channels are always present at the respective systems on the boundary of the box, which is why the circuit on the right-hand side of Fig. \ref{fig:17b} can be claimed to be equivalent to the single transformation on the left-hand side. (The boxes containing  $\mathcal{U}_1$ and  $\mathcal{U}_2$ in this construction are not `bulky' from a temporal perspective as each has a trivial `time span' from input to output, but this is merely due to the simplifying choices we have made. They could be `bulky' if instead of a single transformation on the left-hand side we consider the same transformation composed with identity channels of nontrivial time spans along the boundary systems. One may think that identity channels are also present along the boundary systems on the left-hand side of Fig. \ref{fig:17b}, but they have been suppressed due to being of zero time span and hence entirely trivial.) Using that the middle box on the right-hand side of Fig. \ref{fig:17b} can be further decomposed into two parallel boxes because it contains a tensor product of two operations, $\mathcal{U}_A^{\tilde{A}_I \rightarrow \tilde{A}_O}\otimes\mathcal{I}^{\overline{\tilde{A}}_I\rightarrow \overline{\tilde{A}}_O }$, and the fact that the equivalence in Fig. \ref{fig:17b} must hold for every choice of Alice's unitary and all operations that might be applied by David, Bob, and Charlie (whose correlations fully characterize the process matrix connecting the slots in which these operations are plugged), we conclude the equivalence in Fig. \ref{fig:18}. From Figs. \ref{fig:15}, \ref{fig:17b}, and \ref{fig:18}, it then follows that the the left-hand side of Fig. \ref{fig:18} is equivalent to the right-hand side of Fig. \ref{fig:15}, i.e., the equivalence in Fig. \ref{fig:14} holds for $A_O\equiv \tilde{A}_O$ and $A_I\equiv \tilde{A}_I$ as defined above. Therefore, if we define the system $A_I$ to be the subsystem $\tilde{A}_I$ and $A_O$ to be the subsystem $\tilde{A}_O$ in the circuit on the right-hand side of Fig. \ref{fig:13}, we obtain the equivalence in Fig. \ref{fig:13}. This equivalence is testable by tomography in a way analogous to that explained in Fig. \ref{fig:7}. This completes the proof.

 \section{A generalization}\label{section7} 
 
\begin{figure}
\vspace{0.5 cm}
\begin{center}
\includegraphics[width=7cm]{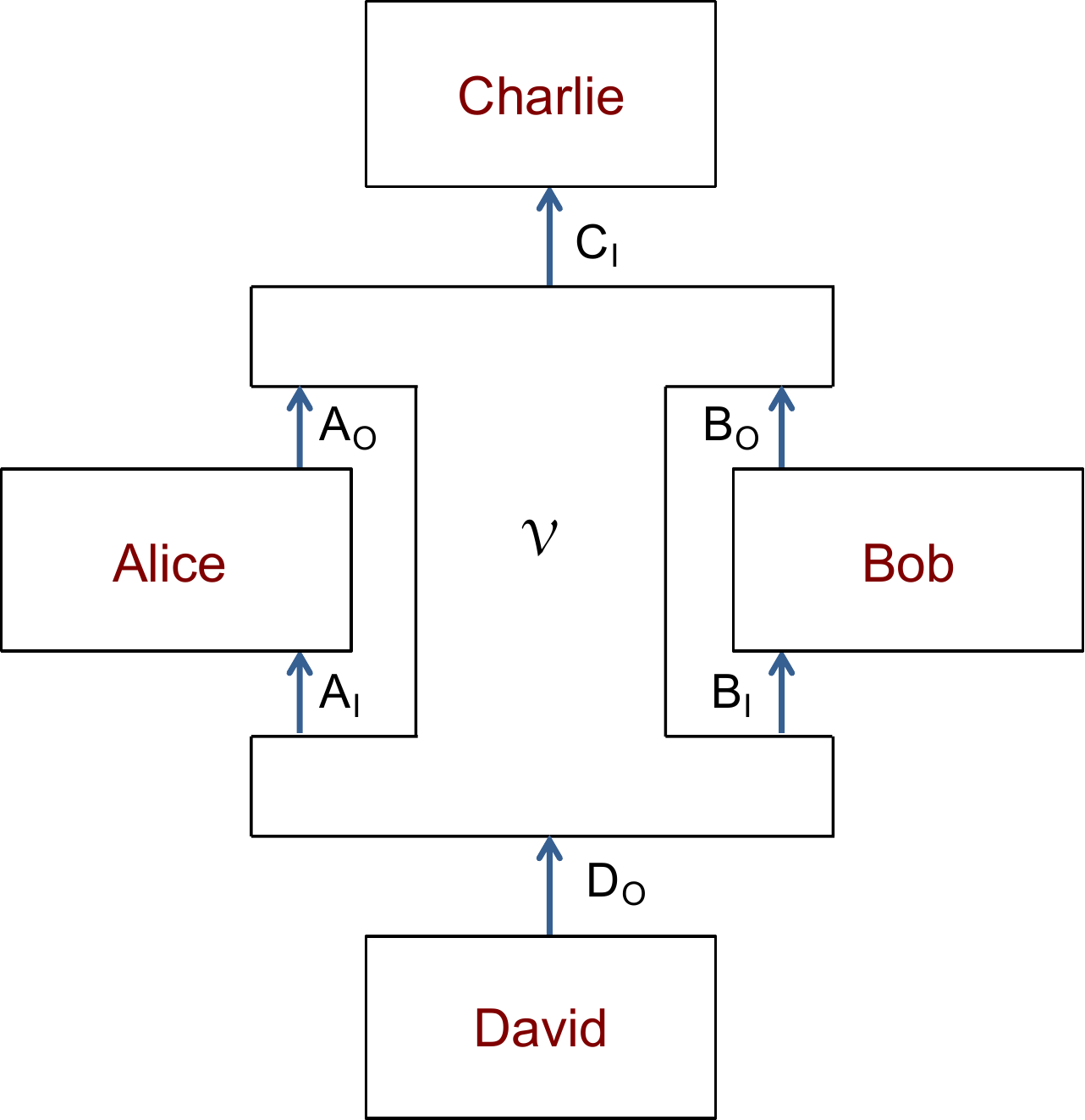}
\end{center}
\vspace{-0.5 cm}
\caption{\textbf{A class of isometric processes that have a realization on time-delocalized subsystems.} We consider the class of four-partite processes of the kind depicted in this figure, which are equivalent to isometric channels that map a subsystem of $D_{O}B_{O}$ isomorphically to $A_{I}$, or a subsystem of $D_{O}A_{O}$ isomorphically to $B_{I}$. We show that they admit realizations on time-delocalized subsystems (see Theorem 2).} \label{fig:19}
\end{figure}

Consider the class of four-partite processes of the kind depicted in Fig. \ref{fig:19}, where the process is equivalent to an isometric channel $\mathcal{V}$ for which the isometry at the Hilbert-space level has the following property. With respect to at least one of the parties Alice and Bob, say Alice, the isometry can be written in the form
\begin{gather}
V^{D_{O}A_{O}B_{O}\rightarrow A_{I}B_{I}C_{I}} = \id^{\tilde{A}_{I} \rightarrow A_{I} }  \otimes V_{2}^{A_{O} \overline{\tilde{A}}_{I} \rightarrow B_IC_I },\label{generalization}
\end{gather}
where $\mathcal{H}^{D_{O}B_{O}} = \mathcal{H}^{\tilde{A}_{I}} \otimes \mathcal{H}^{\overline{\tilde{A}_{I}}} $, and $V_{2}^{A_{O} \overline{\tilde{A}}_{I} \rightarrow B_IC_I }$ is an isometry. Note that $V_{2}^{A_{O} \overline{\tilde{A}}_{I} \rightarrow B_IC_I }$ maps $\mathcal{H}^{A_O} \otimes \mathcal{H}^{ \overline{\tilde{A}}_{I} }$ isomorphically to a subspace of $\mathcal{H}^{B_IC_I}$, and in particular each of the subsystems $\mathcal{H}^{A_O}$ and $ \mathcal{H}^{ \overline{\tilde{A}}_{I} }$ to tensor factors of that subspace. Calling these tensor factors $\mathcal{H}^{\tilde{A}_O}$ and $\mathcal{H}^{\overline{\tilde{A}}_O}$, respectively, we can equivalently write condition \eqref{generalization} as
\begin{gather}
V^{D_{O}A_{O}B_{O}\rightarrow A_{I}B_{I}C_{I}} = \id^{\tilde{A}_{I} \rightarrow A_{I} }  \otimes \id^{A_O \overline{\tilde{A}}_{I} \rightarrow {\tilde{A}_O} {\overline{\tilde{A}}_O} }, \label{generalization2}
\end{gather}
where $\mathcal{H}^{\tilde{A}_O} \otimes \mathcal{H}^{\overline{\tilde{A}}_O}$ is a subspace of $\mathcal{H}^{B_IC_I}$, i.e., 
\begin{gather}
\mathcal{H}^{B_IC_I} = ( \mathcal{H}^{\tilde{A}_O} \otimes \mathcal{H}^{\overline{\tilde{A}}_O}) \oplus \mathcal{H}^K.\label{nontrivialsubsystem}
\end{gather}

Note that there exists an isometric extension of any process, and in particular any bipartite process, which can be obtained by purifying the channel of the process onto Charlie's input (in this case David's output system is trivial), but this purification in general would not have the form \eqref{generalization}.\\

\begin{figure}
	\vspace{0.5 cm}
	\begin{center}
		\includegraphics[width=14cm]{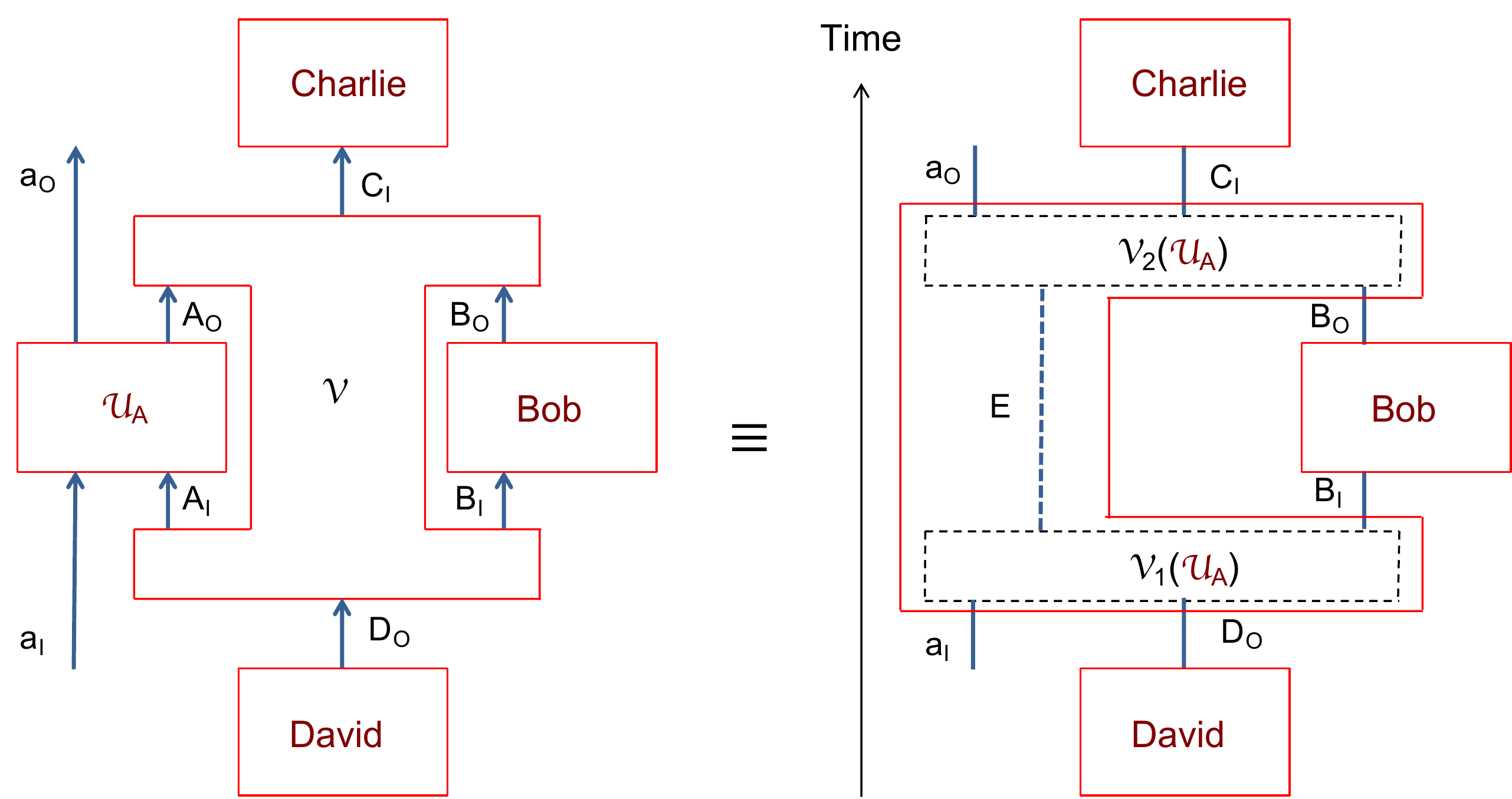}
	\end{center}
	\vspace{-0.5 cm}
	\caption{\textbf{Equivalence of specific circuits and isometric higher-order processes.} According to Theorem 2, there is a subsystem (tensor factor) $A_I$ of the time-local systems $D_OB_O$ depicted on the right-hand side, and a subsystem (tensor factor of a subspace) $A_O$ of the time-local systems $B_IC_I$ depicted on the right-hand side, such that for appropriately chosen isometric channels $\mathcal{V}_1(\mathcal{U}_A)$ and $\mathcal{V}_2(\mathcal{U}_A)$, the circuit on the right-hand side is equivalent to the process on the left-hand side (both understood verifiable box-wise).} \label{fig:20}
\end{figure}

\textbf{Theorem 2 (Realizability of a class of isometric extensions of bipartite processes).} Consider an isometric four-partite process of the kind depicted in Fig. \ref{fig:20}, where the isometry satisfies Eq. \eqref{generalization} (or, equivalently, Eqs. \eqref{generalization2} and \eqref{nontrivialsubsystem}). In agreement with the known quantum mechanics, there exist time-local systems $D_O$, $B_I$, $B_O$, $C_I$, ordered in time in this order, a subsystem (tensor factor) $A_I$ of $D_OB_O$, and a subsystem (tensor factor of a subspace) $A_O$ of $B_IC_I$, such that the circuit-like process depicted on the left-hand side of Fig. \ref{fig:20}, for arbitrary operations of David, Bob, and Charlie (left unspecified in the figure for simplicity), and an arbitrary unitary operation $\mathcal{U}_A^{a_IA_I \rightarrow a_OA_O }$ of Alice (which allows obtaining arbitrary operations in Alice's slot), where $a_I$ and $a_O$ are ancillary input and output systems, admits an experimental implementation on these systems without post-selection. The implementation has a temporal description as on the right-hand side of Fig. \ref{fig:20}, where the isometries $\mathcal{V}_1$ and $\mathcal{V}_2$ are functions of $\mathcal{U}_A$.\\

The proof is similar to the one for the unitary case, so we will not repeat all steps but will only highlight the differences. Here, the channels $\mathcal{U}_1(\mathcal{U}_A)$ and $\mathcal{U}_2(\mathcal{U}_A)$ from the unitary case have been replaced by isometric channels $\mathcal{V}_1(\mathcal{U}_A)$ and $\mathcal{V}_2(\mathcal{U}_A)$, respectively, which is because the composition of Alice's unitary with the isometric process yields a quantum 1-comb that is isometric. (A simple way of seeing this is to note that since the CJ operators of all operations involved are rank-one, so is the operator of the supermap resulting from connecting these operations, and since the latter is equivalent to a quantum comb, which is a channel, the channel can only be isometric.) As before, such isometric channels that realize the minimal Stinespring dilation of the quantum comb (without the need for any auxiliary systems traced out at the end) indeed exist as shown in Ref. \cite{networks}, and they can be seen to be unique up to a joint transformation that is formally equivalent to a change of basis of the intermediary system $E$. 

\begin{figure}
	\vspace{0.5 cm}
	\begin{center}
		\includegraphics[width=13cm]{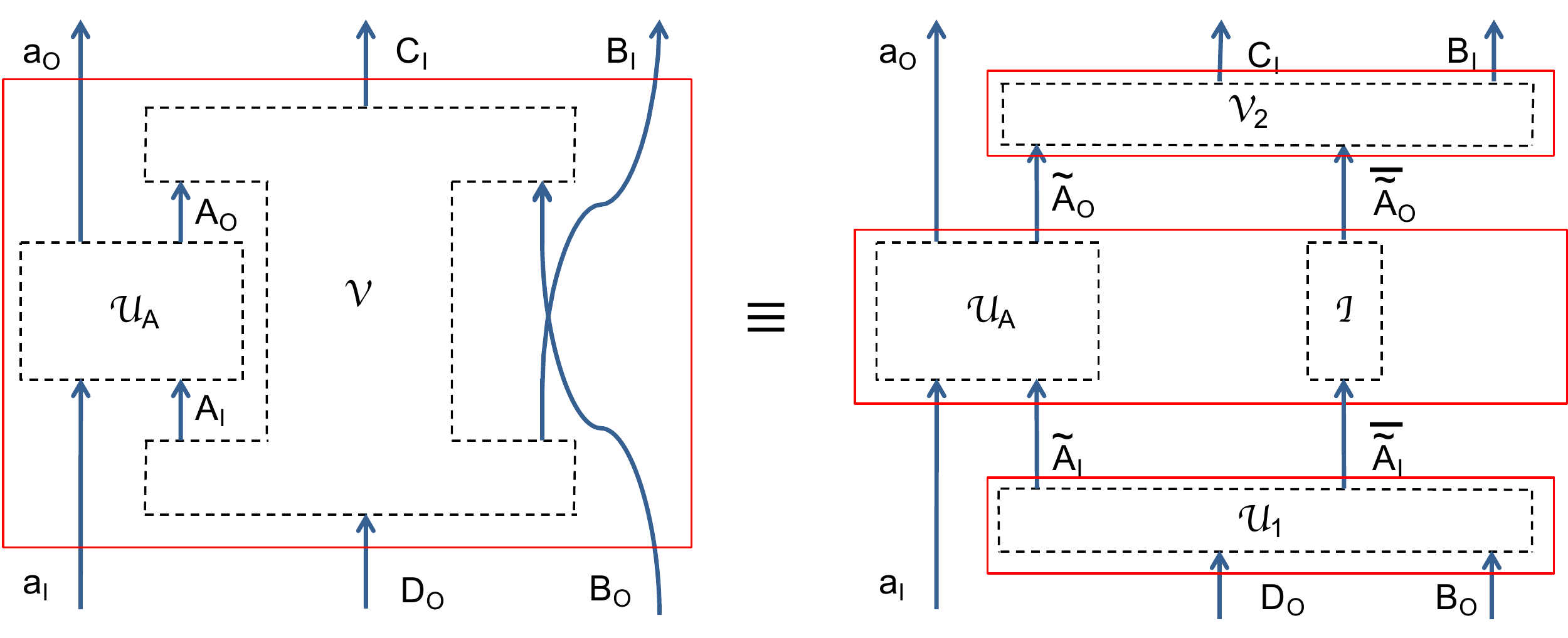}
	\end{center}
	\vspace{-0.5 cm}
	\caption{\textbf{The analogue of Fig. \ref{fig:17b}.} From Eq. \eqref{generalization2}, one can see that the analogue of Fig. \ref{fig:17a} is now Fig. \ref{fig:21}. Here, $\mathcal{U}_1$ is the unitary channel describing the relation between the two different factorizations of the input Hilbert space, and the isometry $\mathcal{V}_2$ describes the embedding of the subspace $\mathcal{H}^{\tilde{A}_O\overline{\tilde{A}}_O}$ of $\mathcal{H}^{B_IC_I} $ into $ \mathcal{H}^{B_IC_I}$. } \label{fig:21}
\end{figure}

From Eq. \eqref{generalization2}, one can directly see that the analogue of Fig. \ref{fig:17a} is now Fig. \ref{fig:21}, where $\mathcal{U}_1$ is the unitary channel describing the relation between the two different factorizations of the input Hilbert space, and the isometry $\mathcal{V}_2$ describes the embedding of the subspace $\mathcal{H}^{\tilde{A}_O\overline{\tilde{A}}_O}$ of $\mathcal{H}^{B_IC_I} $ into $ \mathcal{H}^{B_IC_I}$. Again, these transformations are always present at the respective boundary systems, which is why the single transformation on the left-hand side of Fig. \ref{fig:21} can be claimed to have the nontrivial compositional structure depicted on the right-hand side. (As before, we remark that here we have taken these boundary transformations to be `non-bulky' from a temporal perspective for simplicity, but they could also be `bulky' as explained previously.)

Finally, we note that the isometry of the 1-comb on the right-hand side of Fig. \ref{fig:20}, which by construction equals the transformation on the left-hand side of Fig. \ref{fig:21}, has a form analogous to Eq. \eqref{combfactorization}, 
\begin{gather}
V^{a_{I} D_{O}B_{O}    \rightarrow a_{O}  B_{I}C_{I}  }{(U_A)}= U_A^{a_{I}\tilde{A}_{I}\rightarrow  a_{O}\tilde{A}_{O}}\otimes \id^{\overline{\tilde{A}_{I}} \rightarrow \overline{\tilde{A}_{O}} },
\end{gather}
but with $\tilde{A}_O$ and $\overline{\tilde{A}}_O$ being subsystems in the sense of Eq. \eqref{nontrivialsubsystem}. As before, we define the input and output systems of Alice to be the subsystems $\tilde{A}_I$ and $\tilde{A}_O$ within the circuit on the right-hand side of Fig. \ref{fig:20}. The operation of Alice can be verified experimentally by tomographic operations on the respective subsystems, similarly to all elements of the process. A minor subtlety in comparison to the previous case is the notion of measurement in a tensor factor of a subspace (the subsystem $\tilde{A}_O$ in Eq. \eqref{nontrivialsubsystem}), which can be naturally defined as performing a measurement in the respective factor conditionally on first projecting the full system into the respective subspace (which in this case would succeed with unit probability since the 1-comb on the right-hand side of Fig. \ref{fig:20} always outputs a state in that subspace).  

How big is the class of processes defined here? The four-partite processes are obviously strictly more general than the unitary class discussed in the previous section because they do not have to be unitary and the unitary case is included. However, without additional considerations, it is in principle conceivable that all processes in this class admit unitary extensions and hence they do not describe anything that cannot be achieved with the unitary class. If the latter is not the case, our demonstration that these processes have a realization on suitable time-delocalized subsystems would be a counterexample to the postulate conjectured in Ref. \cite{AraujoPostulate}. An even more radical possibility is that any bipartite process has an extension to a process of this kind, which would imply the realizability of all bipartite processes on suitable time-delocalized subsystems. These questions are left for future investigation.

\section{Discussion}\label{section8}

We have shown that a class of causally nonseparable quantum processes has a physical realization within standard quantum mechanics in terms of operations whose input and output systems are time-delocalized subsystems---a concept that is both mathematically well defined and directly experimentally testable. This result puts on solid grounds the interpretation of recent experimental demonstrations of the quantum SWITCH \cite{Procopio, Rubino1, Rubino2, Goswami} as realizations of this higher-order transformation. A natural addition to this line of demonstrations suggested by the present work is to probe directly the time-delocalized operations of Alice and Bob, e.g., through tomography, as sketched in Fig.~\ref{fig:7}. 

The concept of time-delocalized quantum subsystems and operations introduced here substantially expands the landscape of information-processing structures known to be available within quantum mechanics \cite{networks}, which opens up a new realm of practical possibilities. The irreducible cyclic circuits unveiled in this work are just a glimpse of this landscape, the full scope of which has to be explored through the hierarchy of higher-order supermaps \cite{Perinotti}. 

Apart from understanding the set of physically realizable supermaps, a big open problem is to understand their information processing power. Since those of them incompatible with definite causal order are still part of standard quantum mechanics, the advantages that they can offer over standard causal processes depend on the precise definition of the resources used and their costs. The fact that causally nonseparable processes are more powerful than causally separable ones has been shown by examples of specific computation and communication tasks for which the quantum SWITCH offers advantage \cite{Chiribella12b, Colnaghi, Araujo, Feix, Guerin}. Identifying natural applications for which causal nonseparability would be useful is an important direction for research. 

A specific problem, left unanswered by this work, is whether quantum processes violating causal inequalities have a physical realization. From the circuit perspective emphasized in this paper, such processes are cyclic circuits that give rise to correlations that cannot be simulated by probabilistic mixtures of acyclic circuits or dynamical variants thereof \cite{OG}, even if the systems and operations involved are described by theories more general than quantum theory. In analogy with device-independent applications based on Bell-inequality violations \cite{Brunner}, the existence of such processes could open the road to a new kind of device-independent information-processing applications. The time-delocalized subsystem concept provides a concrete framework in which the question of whether this resource is available in quantum mechanics can be systematically investigated. The present proof that all unitary extensions of bipartite processes, as well as a class of isometric extensions, have a realization on time-delocalized subsystems is a promising step in this direction. As pointed out in Ref. \cite{AraujoPostulate}, there are tripartite processes, such as a deterministic classical process discovered in Ref. \cite{Baumeler3}, which violate causal inequalities and admit a unitary extension. An obvious question is whether and how the proof presented here could be generalized to the case of extensions of multipartite processes \cite{Guerincomment}.
	
One of the results of our analysis of the quantum SWITCH and its implementations is that the operations of Alice and/or Bob are not operations on the target system alone, as sometimes assumed on intuitive grounds, but they spread onto the control qubit as well. While this is consistent with the definition of the quantum SWITCH as a supermap, it may have implications for arguments that treat these operations as operations on the target system \cite{Ebler}. 

Finally, a natural question is whether time-delocalized operations such as those identified here could be seen in a way analogous to standard operations, with the input system being in the past of the output system, with respect to a suitable notion of time, e.g., based on quantum time reference frames. Recent arguments about background-independence in a probabilistic setting \cite{OC2} as well as gedanken proposals for realizations of the quantum SWITCH at the interface of quantum mechanics and gravity \cite{Zych} or with Rindler observers \cite{Dimic} offer a promising route for investigating this subject, as well as the implications that the subsystem picture uncovered in this work may have for quantum gravity.


\acknowledgements{O. O. thanks J. Barrett, \v{C}. Brukner, G. Chiribella, F. Costa, and P. Perinotti for related discussions. The author is grateful to the anonymous reviewers whose comments and questions helped improve the manuscript. Their contributions include spotting a few errors in a previous version of the manuscript, raising the question about the link between time-delocalized subsystems and simulations of the quantum SWITCH with multiple uses of the input operations, and pointing to the need for a better explanation of the argument in Section \ref{section6}, as well as the assumed meaning of `realization'. This work was supported by the Wiener-Anspach Foundation, the UK NQIT Quantum Hub, and the Program of Concerted Research Actions (ARC) of the Université libre de Bruxelles. O. O. is a Research Associate of the Fonds de la Recherche Scientifique (F.R.S.--FNRS).}


\begin{thebibliography}{1}


\bibitem{hardyqg1} L. Hardy, Probability Theories with Dynamic Causal Structure: A New Framework for Quantum Gravity, Preprint at \href{http://arxiv.org/abs/gr-qc/0509120}{$\langle$http://arxiv.org/abs/gr-qc/0509120$\rangle$} (2005).



\bibitem{hardyqg2} L. Hardy, Quantum Gravity Computers: On the Theory of Computation with Indefinite Causal Structure, in \textit{Quantum Reality, Relativistic Causality, and Closing the Epistemic Circle}, The Western Ontario Series in Philosophy of Science, vol. 73 (Springer, Dordrecht, 2009); DOI: \href{https://doi.org/10.1007/978-1-4020-9107-0_21}{https://doi.org/10.1007/978-1-4020-9107-0$\_$21}; Preprint at \href{http://arxiv.org/abs/quant-ph/0701019}{$\langle$http://arxiv.org/abs/quant-ph/0701019$\rangle$} (2007).



\bibitem{Chiribella12}
G. Chiribella, G. M. D'Ariano, P. Perinotti, and B. Valiron, Quantum computations without definite causal structure, \textit{Phys. Rev. A} \textbf{88}, 022318 (2013); DOI: \href{https://doi.org/10.1103/PhysRevA.88.022318}{https://doi.org/10.1103/PhysRevA.88.022318}; Preprint at \href{http://arXiv.org/abs/0912.0195}{$\langle$http://arXiv.org/abs/0912.0195$\rangle$} (2009).

\bibitem{Chiribella08}
G. Chiribella, G. M. D'Ariano, and P. Perinotti, Transforming quantum operations: quantum supermaps, \textit{Europhys. Lett.} \textbf{83}, 30004 (2008); DOI: \href{https://doi.org/10.1209/0295-5075/83/30004}{https://doi.org/10.1209/0295-5075/83/30004}; Preprint at \href{https://arxiv.org/abs/0804.0180}{$\langle$https://arxiv.org/abs/0804.0180$\rangle$} (2008).

\bibitem{networks}
G. Chiribella, G. M. D'Ariano,  and P. Perinotti, Theoretical framework for quantum networks,
\textit{Phys. Rev. A} \textbf{80}, 022339 (2009); DOI: \href{https://doi.org/10.1103/PhysRevA.80.022339}{https://doi.org/10.1103/PhysRevA.80.022339}; Preprint at \href{http://arxiv.org/abs/0904.4483}{$\langle$http://arxiv.org/abs/0904.4483$\rangle$} (2009).

\bibitem{onDeutsch} The idea of quantum computation beyond causal circuits was notably first considered by Deutsch \cite{Deutsch} who studied modifications of quantum theory in the vicinity of closed timelike curves, which led to the study of the computational power of such models (see Ref. \cite{Aaronson}). Although linked to time travel in a different sense \cite{Chiribella12}, the quantum SWITCH is motivated by the idea of `quantum superpositions of different causal structures' as opposed to classically definite backgrounds with timelike cycles. 

\bibitem{Deutsch} D. Deutsch, Quantum mechanics near closed timelike lines, \textit{Phys. Rev. D} \textbf{44}, 3197 (1991); DOI: \href{https://doi.org/10.1103/PhysRevD.44.3197}{https://doi.org/10.1103/PhysRevD.44.3197}.

\bibitem{Aaronson} S. Aaaronson and J. Watrous, Closed timelike curves make quantum and classical computing equivalent, \textit{Proc. R. Soc. A} \textbf{465}, 631-647 (2009); DOI: \href{https://doi.org/10.1098/rspa.2008.0350}{https://doi.org/10.1098/rspa.2008.0350}; Preprint at \href{https://arxiv.org/abs/0808.2669}{$\langle$https://arxiv.org/abs/0808.2669$\rangle$} (2008).

\bibitem{OCB} O. Oreshkov, F. Costa, and \v{C}. Brukner, Quantum correlations with no causal order, {\em Nat. Commun.} \textbf{3}, 1092 (2012); DOI: \href{https://doi.org/10.1038/ncomms2076}{https://doi.org/10.1038/ncomms2076}; Preprint at \href{http://arxiv.org/abs/1105.4464}{$\langle$http://arxiv.org/abs/1105.4464$\rangle$} (2011).


\bibitem{OG} O. Oreshkov and C. Giarmatzi, Causal and causally separable processes, \textit{New J. Phys.} \textbf{18}, 093020 (2016); DOI: \href{https://doi.org/10.1088/1367-2630/18/9/093020}{https://doi.org/10.1088/1367-2630/18/9/093020}; Preprint at \href{http://arxiv.org/abs/1506.05449}{$\langle$http://arxiv.org/abs/1506.05449$\rangle$} (2015). 


\bibitem{AraujoWitness} M. Ara\'{u}jo, C. Branciard, F. Costa, A. Feix, C. Giarmatzi, and \v{C}. Brukner, Witnessing causal nonseparability, \textit{New J. Phys.} \textbf{17}, 102001 (2015); DOI: \href{https://doi.org/10.1088/1367-2630/17/10/102001}{https://doi.org/10.1088/1367-2630/17/10/102001}; Preprint at \href{http://arxiv.org/abs/1506.03776}{$\langle$http://arxiv.org/abs/1506.03776$\rangle$} (2015).

\bibitem{Wechs} J. Wechs, A. A. Abbott, and C. Branciard, On the definition and characterisation of multipartite causal (non)separability, \textit{New J. Phys.} \textbf{21}, 013027 (2019); DOI: \href{https://doi.org/10.1088/1367-2630/aaf352}{https://doi.org/10.1088/1367-2630/aaf352}; Preprint at \href{https://arxiv.org/abs/1807.10557}{$\langle$https://arxiv.org/abs/1807.10557$\rangle$} (2018).

\bibitem{Baumeler1}   \"{A}. Baumeler and S. Wolf, Perfect signaling among three parties violating predefined causal order, Proceedings of International Symposium on Information Theory (ISIT) 2014, 526-530, 2014; DOI: \href{https://doi.org/10.1109/ISIT.2014.6874888}{https://doi.org/10.1109/ISIT.2014.6874888}; Preprint at \href{http://arxiv.org/abs/1312.5916}{$\langle$http://arxiv.org/abs/1312.5916$\rangle$} (2013).

\bibitem{Baumeler2} \"{A}. Baumeler, A. Feix, and S. Wolf, Maximal incompatibility of locally classical behavior and global causal order in multi-party scenarios, \textit{Phys. Rev. A} \textbf{90}, 042106 (2014); DOI: \href{https://doi.org/10.1103/PhysRevA.90.042106}{https://doi.org/10.1103/PhysRevA.90.042106}; Preprint at \href{http://arxiv.org/abs/1403.7333}{$\langle$http://arxiv.org/abs/1403.7333$\rangle$} (2014).



\bibitem{Baumeler3} \"{A}. Baumeler and S. Wolf, The space of logically consistent classical processes without causal order, \textit{New J. Phys.} \textbf{18}, 013036 (2016); DOI: \href{https://doi.org/10.1088/1367-2630/18/1/013036}{https://doi.org/10.1088/1367-2630/18/1/013036}; Preprint at \href{http://arxiv.org/abs/1507.01714}{$\langle$http://arxiv.org/abs/1507.01714$\rangle$} (2015).


\bibitem{branciard} C. Branciard, M. Ara\'{u}jo, A. Feix, F. Costa, and \v{C}. Brukner, The simplest causal inequalities and their violation, \textit{New J. Phys.} \textbf{18}, 013008 (2016); DOI: \href{https://doi.org/10.1088/1367-2630/18/1/013008}{https://doi.org/10.1088/1367-2630/18/1/013008}; Preprint at \href{http://arxiv.org/abs/1508.01704}{$\langle$http://arxiv.org/abs/1508.01704$\rangle$} (2015).


\bibitem{Bhattacharya} S. S. Bhattacharya and M. Banik, Biased Non-Causal Game, Preprint at \href{http://arxiv.org/abs/1509.02721}{$\langle$http://arxiv.org/abs/1509.02721$\rangle$} (2015).  

\bibitem{Feix2} A. Feix, M. Ara\'{u}jo, and \v{C}. Brukner, Causally nonseparable processes admitting a causal model, \textit{New J. Phys.} \textbf{18}, 083040 (2016); DOI: \href{https://doi.org/10.1088/1367-2630/18/8/083040}{https://doi.org/10.1088/1367-2630/18/8/083040}; Preprint at \href{http://arxiv.org/abs/1604.03391}{$\langle$http://arxiv.org/abs/1604.03391$\rangle$} (2016).

\bibitem{Abbott} A. A. Abbott, C. Giarmatzi, F. Costa, and C. Branciard, Multipartite Causal Correlations: Polytopes and Inequalities, \textit{Phys. Rev. A} \textbf{94}, 032131 (2016); DOI: \href{https://doi.org/10.1103/PhysRevA.94.032131}{https://doi.org/10.1103/PhysRevA.94.032131}; Preprint at \href{http://arxiv.org/abs/1608.01528}{$\langle$http://arxiv.org/abs/1608.01528$\rangle$} (2016).


\bibitem{Miklin} N. Miklin, A. A. Abbott, C. Branciard, R. Chaves, C. Budroni, The entropic approach to causal correlations, \textit{New J. Phys.} \textbf{19}, 113041 (2017); DOI: \href{https://doi.org/10.1088/1367-2630/aa8f9f}{https://doi.org/10.1088/1367-2630/aa8f9f}; Preprint at \href{http://arxiv.org/abs/1706.10270}{$\langle$http://arxiv.org/abs/1706.10270$\rangle$} (2017).





\bibitem{Bell} J. S. Bell, On the Einstein Podolsky Rosen Paradox,
\textit{Physics} \textbf{1}, 3, 195-200 (1964); DOI: \href{https://doi.org/10.1103/PhysicsPhysiqueFizika.1.195}{https://doi.org/10.1103/PhysicsPhysiqueFizika.1.195}.



\bibitem{OC2} O. Oreshkov and N. J. Cerf, Operational quantum theory without predefined time, \textit{New J. Phys.} \textbf{18}, 073037 (2016); DOI: \href{https://doi.org/10.1088/1367-2630/18/7/073037}{https://doi.org/10.1088/1367-2630/18/7/073037}; Preprint at \href{http://arxiv.org/abs/1406.3829}{$\langle$http://arxiv.org/abs/1406.3829$\rangle$} (2014).

\bibitem{Silvaetal} R. Silva, Y. Guryanova, A. J. Short, P. Skrzypczyk, N. Brunner, and S. Popescu, Connecting processes with indefinite causal order and multi-time quantum states, \textit{New J. Phys.} \textbf{19} , 103022 (2017); DOI: \href{https://doi.org/10.1088/1367-2630/aa84fe}{https://doi.org/10.1088/1367-2630/aa84fe}; Preprint at \href{http://arxiv.org/abs/1701.08638}{$\langle$http://arxiv.org/abs/1701.08638$\rangle$} (2017).


\bibitem{AraujoCTC} M. Ara\'{u}jo,  P. A. Gu\'{e}rin, and \"{A}. Baumeler, Quantum computation with indefinite causal structures, \textit{Phys. Rev. A} \textbf{96}, 052315 (2017); DOI: \href{https://doi.org/10.1103/PhysRevA.96.052315}{https://doi.org/10.1103/PhysRevA.96.052315}; Preprint at \href{https://arxiv.org/abs/1706.09854}{$\langle$https://arxiv.org/abs/1706.09854$\rangle$} (2017).


\bibitem{Milz} S. Milz, F. A. Pollock, T. P. Le, G. Chiribella, and K. Modi, Entanglement, non-Markovianity, and causal non-separability, \textit{New J. Phys.} \textbf{20}, 033033 (2018); DOI: \href{https://doi.org/10.1088/1367-2630/aaafee}{https://doi.org/10.1088/1367-2630/aaafee}; Preprint at \href{http://arxiv.org/abs/1711.04065}{$\langle$http://arxiv.org/abs/1711.04065$\rangle$} (2017).




\bibitem{Chiribella12b} G. Chiribella, Perfect discrimination of no-signalling channels via quantum superposition of causal structures, {\em Phys. Rev. A} \textbf{86}, 040301 (2012); DOI: \href{https://doi.org/10.1103/PhysRevA.86.040301}{https://doi.org/10.1103/PhysRevA.86.040301}; Preprint at \href{http://arxiv.org/abs/1109.5154}{$\langle$http://arxiv.org/abs/1109.5154$\rangle$} (2011).



\bibitem{BranciardWitness} C. Branciard, Witnesses of causal nonseparability: an introduction and a few case studies, \textit{Sci. Rep.} \textbf{6}, 26018 (2016); DOI: \href{https://doi.org/10.1038/srep26018}{https://doi.org/10.1038/srep26018}; Preprint at \href{http://arxiv.org/abs/1603.00043}{$\langle$http://arxiv.org/abs/1603.00043$\rangle$} (2016).



\bibitem{Colnaghi} T. Colnaghi, G. M. D'Ariano, P. Perinotti, and S. Facchini, Quantum computation with programmable connections between gates, \textit{Phys. Lett. A} \textbf{376}, 2940 - 2943 (2012); DOI: \href{https://doi.org/10.1016/j.physleta.2012.08.028}{https://doi.org/10.1016/j.physleta.2012.08.028}; Preprint at \href{http://arxiv.org/abs/1109.5987}{$\langle$http://arxiv.org/abs/1109.5987$\rangle$} (2011).



\bibitem{Araujo} M. Ara\'{u}jo, F. Costa, and \v{C}. Brukner, Computational advantage from quantum-controlled ordering of gates, \textit{Phys. Rev. Lett.} \textbf{113}, 250402 (2014); DOI: \href{https://doi.org/10.1103/PhysRevLett.113.250402}{https://doi.org/10.1103/PhysRevLett.113.250402}; Preprint at \href{https://arxiv.org/abs/1401.8127}{$\langle$https://arxiv.org/abs/1401.8127$\rangle$} (2014).

\bibitem{Feix} A. Feix, M. Ara\'{u}jo, and \v{C}. Brukner, Quantum superposition of the order of parties as a communication resource, \textit{Phys. Rev. A} \textbf{92}, 052326 (2015); DOI: \href{https://doi.org/10.1103/PhysRevA.92.052326}{https://doi.org/10.1103/PhysRevA.92.052326}; Preprint at \href{https://arxiv.org/abs/1508.07840}{$\langle$https://arxiv.org/abs/1508.07840$\rangle$} (2016).  


\bibitem{Guerin} P. A. Gu\'{e}rin, A. Feix, M. Ara\'{u}jo, and \v{C}. Brukner, Exponential Communication Complexity Advantage from Quantum Superposition of the Direction of Communication, \textit{Phys. Rev. Lett.} \textbf{117}, 100502 (2016); DOI: \href{https://doi.org/10.1103/PhysRevLett.117.100502}{https://doi.org/10.1103/PhysRevLett.117.100502}; Preprint at \href{http://arxiv.org/abs/1605.07372}{$\langle$http://arxiv.org/abs/1605.07372$\rangle$} (2016).
 

\bibitem{Friis1} N. Friis, V. Dunjko, W. D\"{u}r, and H. J. Briegel, Implementing quantum control for unknown subroutines, \textit{Phys. Rev. A} \textbf{89}, 030303(R) (2014); DOI: \href{https://doi.org/10.1103/PhysRevA.89.030303}{https://doi.org/10.1103/PhysRevA.89.030303}; Preprint at \href{https://arxiv.org/abs/1401.8128}{$\langle$https://arxiv.org/abs/1401.8128$\rangle$} (2014).


\bibitem{Procopio} L. M. Procopio \textit{et al.}, Experimental Superposition of Orders of Quantum Gates,  \textit{Nat. Commun.} \textbf{6}, 7913 (2015); DOI: \href{https://doi.org/10.1038/ncomms8913}{https://doi.org/10.1038/ncomms8913}; Preprint at \href{http://arxiv.org/abs/1412.4006}{$\langle$http://arxiv.org/abs/1412.4006$\rangle$} (2014). 

\bibitem{Friis2} N. Friis, A. A. Melnikov, G. Kirchmair, and H. J. Briegel, Coherent controlization using superconducting qubits, \textit{Sci. Rep.} \textbf{5}, 18036 (2015); DOI: \href{https://doi.org/10.1038/srep18036}{https://doi.org/10.1038/srep18036}; Preprint at \href{https://arxiv.org/abs/1508.00447}{$\langle$https://arxiv.org/abs/1508.00447$\rangle$} (2015).


\bibitem{Rubino1} G. Rubino, L. A. Rozema, A. Feix, M. Ara\'{u}jo, J. M. Zeuner, L. M. Procopio, \v{C}. Brukner, and P. Walther, Experimental Verification of an Indefinite Causal Order, \textit{Sci. Adv.} \textbf{3}, e1602589 (2017); DOI: \href{https://doi.org/10.1126/sciadv.1602589}{https://doi.org/10.1126/sciadv.1602589}; Preprint at \href{http://arxiv.org/abs/1608.01683}{$\langle$http://arxiv.org/abs/1608.01683$\rangle$} (2016).

\bibitem{Rubino2} G. Rubino, L. A. Rozema, F. Massa, M. Ara\'{u}jo, M. Zych, \v{C}. Brukner, and P. Walther, Experimental Entanglement of Temporal Orders, Preprint at \href{http://arxiv.org/abs/1712.06884}{$\langle$http://arxiv.org/abs/1712.06884$\rangle$} (2017). 

\bibitem{Goswami} K. Goswami, C. Giarmatzi, M. Kewming, F. Costa, C. Branciard, J. Romero, and A. G. White, Indefinite Causal Order in a Quantum Switch, \textit{Phys. Rev. Lett.} \textbf{121}, 090503 (2018); DOI: \href{https://doi.org/10.1103/PhysRevLett.121.090503}{https://doi.org/10.1103/PhysRevLett.121.090503}; Preprint at \href{https://arxiv.org/abs/1803.04302}{$\langle$https://arxiv.org/abs/1803.04302$\rangle$} (2018). 


\bibitem{Viola} L. Viola, E. Knill, and R. Laflamme, Constructing Qubits in Physical Systems, \textit{J. Phys. A} \textbf{34}, 7067 (2001); DOI \href{https://doi.org/10.1088/0305-4470/34/35/331}{https://doi.org/10.1088/0305-4470/34/35/331}; Preprint at \href{http://arxiv.org/abs/quant-ph/0101090}{$\langle$http://arxiv.org/abs/quant-ph/0101090$\rangle$} (2001).  

\bibitem{Knill} E. Knill, Protected realizations of quantum information, \textit{Phys. Rev. A} \textbf{74}, 042301 (2006); DOI: \href{https://doi.org/10.1103/PhysRevA.74.042301}{https://doi.org/10.1103/PhysRevA.74.042301}; Preprint at \href{http://arxiv.org/abs/quant-ph/0603252}{$\langle$http://arxiv.org/abs/quant-ph/0603252$\rangle$} (2006). 

\bibitem{KribsSpekkens} D. W. Kribs and R. W. Spekkens, Quantum Error Correcting Subsystems are Unitarily Recoverable Subsystems, \textit{Phys. Rev. A} \textbf{74}, 042329 (2006); DOI: \href{https://doi.org/10.1103/PhysRevA.74.042329}{https://doi.org/10.1103/PhysRevA.74.042329}; Preprint at \href{https://arxiv.org/abs/quant-ph/0608045}{$\langle$https://arxiv.org/abs/quant-ph/0608045$\rangle$} (2006).

\bibitem{Zanardi} P. Zanardi, Virtual Quantum Subsystems, \textit{Phys. Rev. Lett.} \textbf{87}, 077901 (2001); DOI: \href{https://doi.org/10.1103/PhysRevLett.87.077901}{https://doi.org/10.1103/PhysRevLett.87.077901}; Preprint at \href{http://arxiv.org/abs/quant-ph/0103030}{$\langle$http://arxiv.org/abs/quant-ph/0103030$\rangle$} (2001).

\bibitem{ZLL} P. Zanardi, D. Lidar, and S. Lloyd, Quantum Tensor Product Structures are Observable Induced, \textit{Phys. Rev. Lett.} \textbf{92}, 060402 (2004); DOI: \href{https://doi.org/10.1103/PhysRevLett.92.060402}{https://doi.org/10.1103/PhysRevLett.92.060402}; Preprint at \href{http://arxiv.org/abs/quant-ph/0308043}{$\langle$http://arxiv.org/abs/quant-ph/0308043$\rangle$} (2003).

\bibitem{AraujoPostulate} M. Ara\'{u}jo, Adrien Feix, Miguel Navascu\'{e}s, and \v{C}aslav Brukner, A purification postulate for quantum mechanics with indefinite causal order, \textit{Quantum} \textbf{1}, 10 (2017); DOI: \href{https://doi.org/10.22331/q-2017-04-26-10}{https://doi.org/10.22331/q-2017-04-26-10}; Preprint at \href{http://arxiv.org/abs/1611.08535}{$\langle$http://arxiv.org/abs/1611.08535$\rangle$} (2016).














\bibitem{jam} A. Jamio{\l}kowski, Linear transformations which preserve trace and positive semidefiniteness of operators, \textit{Rep. Math. Phys.} \textbf{3}, 4, 275-278 (1972); DOI \href{https://doi.org/10.1016/0034-4877(72)90011-0}{https://doi.org/10.1016/0034-4877(72)90011-0}.

\bibitem{choi} M.-D. Choi, Completely positive linear maps on complex matrices, \textit{Lin. Alg. Appl.} \textbf{10}, 285-290 (1975); DOI: \href{https://doi.org/10.1016/0024-3795(75)90075-0}{https://doi.org/10.1016/0024-3795(75)90075-0}.





\bibitem{OC1} O. Oreshkov and N. J. Cerf, Operational formulation of time reversal in quantum theory, \textit{Nature Phys.} \textbf{11}, 853-858 (2015); DOI: \href{https://doi.org/10.1038/nphys3414}{https://doi.org/10.1038/nphys3414}; Preprint at  \href{http://arxiv.org/abs/1507.07745}{$\langle$http://arxiv.org/abs/1507.07745$\rangle$} (2015).








\bibitem{Perinotti} P. Perinotti, Causal Structures and the Classification of Higher Order Quantum Computations, in \textit{Time in physics}, R. Renner and S. Stupar (eds), Tutorials, Schools, and Workshops in the Mathematical Sciences, (Birkh\"{a}user, Cham, 2017); DOI: \href{https://doi.org/10.1007/978-3-319-68655-4_7}{https://doi.org/10.1007/978-3-319-68655-4$\_$7}; Preprint at \href{http://arxiv.org/abs/1612.05099}{$\langle$http://arxiv.org/abs/1612.05099$\rangle$} (2016).

\bibitem{Kissinger} A. Kissinger and S. Uijlen, A categorical semantics for causal structure, \textit{Logical Methods in Computer Science}, Volume 15, Issue 3 (August 9, 2019), lmcs:5681; DOI: \href{https://doi.org/10.23638/LMCS-15(3:15)2019}{https://doi.org/10.23638/LMCS-15(3:15)2019}; Preprint at \href{http://arxiv.org/abs/1701.04732}{$\langle$http://arxiv.org/abs/1701.04732$\rangle$} (2017).


\bibitem{NielsenChuang} M. A. Nielsen and I. L. Chuang, \textit{Quantum computation and quantum information}, (Cambridge University Press, Cambridge,
2000); DOI \href{https://doi.org/10.1017/CBO9780511976667}{https://doi.org/10.1017/CBO9780511976667}.

\bibitem{Esteban} E. Castro-Ruiz, F. Giacomini, and \v{C}aslav Brukner, Dynamics of Quantum Causal Structures, \textit{Phys. Rev. X} \textbf{8}, 011047 (2018); DOI: \href{https://doi.org/10.1103/PhysRevX.8.011047}{https://doi.org/10.1103/PhysRevX.8.011047}; Preprint at \href{https://arxiv.org/abs/1710.03139}{$\langle$https://arxiv.org/abs/1710.03139$\rangle$} (2017).

\bibitem{no-go} Strictly speaking, in Ref. \cite{Chiribella12} it was shown that if there exists a realization of the quantum SWITCH such that Alice's operation is in the past of Bob's operation so that the output ancilla of Alice could be connected to the input ancilla of Bob, this would allow deterministic transmission of information back in time. In the realization discussed here, this condition is not satisfied---the ancillary systems of Alice and Bob cannot be connected to each other as they occupy space-like separated regions. Nevertheless, the full experiment still has the structure of a circuit with a `timelike' cycle, albeit not permitting deterministic time travel, as any quantum process matrix is equivalent to a channel from the output systems of all parties to their input systems \cite{OCB}. 

\bibitem{SWAP} More precisely, $\textrm{C-SWAP}^{XYZ} = |0\rangle\langle 0|^{X}\otimes  \id^{YZ} + |1\rangle\langle 1|^{X}\otimes \textrm{SWAP}^{YZ}$, where $\textrm{SWAP}^{YZ}$ is the SWAP operator on $Y$ and $Z$, which can be defined as follows. Consider two systems $Y$ and $Z$ with Hilbert spaces of the same dimension and a linear isomorphism between the states in these Hilbert spaces. An arbitrary vector in the joint system $YZ$ can be written in the form $|\psi\rangle^{YZ} = \sum_{i,j} \psi_{ij}|i\rangle^Y |j\rangle^Z$, where $\{|i\rangle^Y\}$ are orthonormal bases for $Y$ and $Z$, respectively. The action of the operator $\textrm{SWAP}^{YZ}$ on the vector $|\psi\rangle^{YZ}$ is then given by $\textrm{SWAP}^{YZ} |\psi\rangle^{YZ} = \sum_{i,j} \psi_{ij}|j\rangle^Y |i\rangle^Z$. 




\bibitem{endnoteX} Of course, if during the working of the device, an adversary turns on unwanted interactions, such as a Hamiltonian on the control qubit that is not diagonal in the logical basis, this could prevent the device from implementing the correct operation on the systems of interest. But this is the case for any physical device implementing an operation, irrespectively of whether the operation is localized or delocalized in time. 

\bibitem{CyclicProof} Very recently, after the submission of this paper, the author and colleagues J. Barrett and R. Lorenz showed via different methods that all bipartite processes that are unitarily extendible are causally separable, and hence their unitary extensions are variations of the quantum SWITCH (in preparation). Nevertheless, we believe that the proof of realizability presented here has a particular value since it is based on a different idea that could have wider applications. In particular, it provides the basis for the generalization in Sec. \ref{section7}, and might be useful in the search for realizations of more complicated unitary processes. 

\bibitem{Stinespring} W. F. Stinespring, Positive functions on \textit{C*}-algebras, \textit{Proc. Amer. Math. Soc.} \textbf{6}, 211 (1955); DOI: \href{https://doi.org/10.2307/2032342}{https://doi.org/10.2307/2032342}.

\bibitem{isomorphic} Throughout this paper, when we speak about isomorphic mapping between two Hilbert spaces, we understand isometric isomorphism. 


\bibitem{Brunner} N. Brunner, D. Cavalcanti, S. Pironio, V. Scarani, and S. Wehner, Bell nonlocality, \textit{Rev. Mod. Phys.} \textbf{86}, 419 (2014); DOI: \href{https://doi.org/10.1103/RevModPhys.86.419}{https://doi.org/10.1103/RevModPhys.86.419}; Preprint at \href{http://arxiv.org/abs/1303.2849}{$\langle$http://arxiv.org/abs/1303.2849$\rangle$} (2013).


\bibitem{Guerincomment} After this paper appeared, a subsequent paper \cite{AllardGuerin} claimed to show that all unitary processes admit a representation on time-delocalized subsystems. However, this claim is based on a misunderstanding of the concept of time-delocalized subsystems. The proof claimed in \cite{AllardGuerin} amounts to the observation (discussed in this paper) that if we have a unitary process, the unitary maps isomorphically the output system of any one party, say Alice, onto a subsystem of the input systems of the rest of the parties, and similarly maps a subsystem of the output systems of the rest of the parties onto the input system of Alice. This by itself does not imply that we can associate the input and output systems of Alice with time-delocalized subsystems (which are subsystems of tensor products of Hilbert spaces associated with concrete physical systems at concrete times).

\bibitem{AllardGuerin} P. Allard Guérin and \v{C}. Brukner, Observer-dependent locality of quantum events, \textit{New J. Phys.} \textbf{20}, 103031 (2018); DOI: \href{https://doi.org/10.1088/1367-2630/aae742}{https://doi.org/10.1088/1367-2630/aae742}; Preprint at \href{https://arxiv.org/abs/1805.12429}{$\langle$https://arxiv.org/abs/1805.12429$\rangle$} (2018).   

\bibitem{Ebler} D. Ebler, S. Salek, and G. Chiribella, Enhanced Communication with the Assistance of Indefinite Causal Order, \textit{Phys. Rev. Lett.} {\bf 120}, 120502 (2018); DOI: \href{https://doi.org/10.1103/PhysRevLett.120.120502}{https://doi.org/10.1103/PhysRevLett.120.120502}; Preprint at \href{http://arxiv.org/abs/1711.10165}{$\langle$http://arxiv.org/abs/1711.10165$\rangle$} (2017).

\bibitem{Zych} M. Zych, F. Costa, I. Pikovski, and \v{C}aslav Brukner, Bell's Theorem for Temporal Order, \textit{Nat. Commun.} \textbf{10}, 3772 (2019); DOI: \href{https://doi.org/10.1038/s41467-019-11579-x}{https://doi.org/10.1038/s41467-019-11579-x}; Preprint at \href{http://arxiv.org/abs/1708.00248}{$\langle$http://arxiv.org/abs/1708.00248$\rangle$} (2017).

\bibitem{Dimic} A. Dimi\'{c}, M. Milivojevi\'{c}, D. Go\v{c}anin, and \v{C}aslav Brukner, Simulating spacetime with indefinite causal order \textit{via} Rindler observers, Preprint at \href{http://arxiv.org/abs/1712.02689}{$\langle$http://arxiv.org/abs/1712.02689$\rangle$} (2017).














































































\end{thebibliography}
\end{document}